\documentclass{ws-ijmpe}
\usepackage[super,compress]{cite}
\usepackage{epstopdf}
\usepackage{xcolor}

\usepackage{graphicx}
\usepackage{dcolumn}
\usepackage{bm}
\usepackage{amssymb}
\usepackage{xspace}

\usepackage{hyperref}
\usepackage{amsmath}

\newcommand{\pp}{\ensuremath{\rm pp}\xspace}

\newcommand{\raa}{\ensuremath{R_{\rm AA}}\xspace}

\newcommand{\pt}{\ensuremath{p_{\rm{T}}}\xspace}

\newcommand{\pbpb}{Pb--Pb\xspace}
\newcommand{\auau}{Au--Au\xspace}
\newcommand{\cucu}{Cu--Cu\xspace}
\newcommand{\ppb}{p--Pb\xspace}
\newcommand{\pa}{p--A\xspace}

\newcommand{\snnt}[1]{\ensuremath{\sqrt{s_{\rm NN}} = #1 \text{\,TeV}}\xspace}

\newcommand{\sppt}[1]{\ensuremath{\sqrt{s} = #1 \text{\,TeV}}\xspace}

\newcommand{\Ntrig}{\ensuremath{N^\mathrm{AA}_{\rm{trig}}}}
\newcommand{\dNjetdpT}{ \frac{ {\rm d^2} \ensuremath{N^{\rm AA}_{\rm jet}}} {\mathrm{d}\pTjetch\mathrm{d}\eta_\mathrm{jet}}}

\newcommand{\pTjetch}{\ensuremath{p_\mathrm{T,jet}^\mathrm{ch}}}
\newcommand{\pTtrig}{\ensuremath{p_{\mathrm{T,trig}}}}

\newcommand{\AAtohjet}{\mathrm{AA}\rightarrow\rm{h}+{jet}+X}
\newcommand{\AAtoh}{\mathrm{AA}\rightarrow\rm{h}+X}
\newcommand{\pTh}{\ensuremath{p_{\mathrm{T,h}}}}
\newcommand{\Drecoil}{\ensuremath{\Delta_\mathrm{recoil}}}

\newcommand{\TTSig}{\ensuremath{\mathrm{TT}_{\mathrm{Sig}}}}
\newcommand{\TTRef}{\ensuremath{\mathrm{TT}_{\mathrm{Ref}}}}
\newcommand{\pT}{\ensuremath{p_{\rm{T}}}}
\newcommand{\RdR}{\mbox{$R_{\Delta R}$}}

\begin{document}

\markboth{Bala, Bautista, Biel\v{c}\'{i}kov\'a and Ortiz}{Heavy-ion physics at the LHC}

\catchline{}{}{}{}{}


\title{Heavy-ion physics at the LHC: Review of Run I results}

\author{Renu Bala}

\address{Department of Physics, University of Jammu, Jammu, India\\
Renu.Bala@cern.ch}

\author{Irais Bautista}

\address{CONACYT Research Fellow- Facultad de Ciencias F\'isico Matem\'aticas, Benem\'erita Universidad Aut\'onoma de Puebla, 1152, M\'exico.\\
irbautis@cern.ch}

\author{Jana Biel\v{c}\'{i}kov\'a}

\address{Nuclear Physics Institute, Czech Academy of Sciences, 250 68 \v{R}e\v{z}, Czech Republic \\
jana.bielcikova@ujf.cas.cz}

\author{Antonio Ortiz}

\address{Instituto de Ciencias Nucleares, UNAM, Ciudad de M\'exico, 04510, M\'exico\\
antonio.ortiz@nucleares.unam.mx}

\maketitle

\begin{history}
\received{Day Month Year}
\revised{Day Month Year}
\end{history}

\begin{abstract}
In this work we review what we consider are, some of the most relevant results of heavy-ion physics at the LHC. This paper is not intended to cover all the many important results of the experiments, instead we present a brief overview of the current status on the characterization of the hot and dense QCD medium produced in the heavy-ion collisions. Recent exciting results which are still under debate are discussed too, leading to intriguing questions like whether we have a real or fake QGP formation in small systems.

\end{abstract}

\keywords{QGP; LHC; QCD.}

\ccode{PACS numbers:}


\section{Introduction}
\label{sec:1}

Quantum-ChromoDynamics (QCD) is the theory of the strong interaction, which describes the binding of quarks by gluons to form hadrons such as neutrons and protons. The QCD coupling constant ($\alpha_{\rm s}$) is a strongly varying function of energy, e.g., at low momentum transfers (low-$Q^{2}$) it becomes large and perturbative QCD (pQCD) cannot be applied. From the theory side, the non-perturbative QCD regime has not been explored as much as pQCD. Though, this is the regime where a new phase of QCD matter, the Quark-Gluon Plasma (QGP), can be studied. From Lattice Quantum-ChromoDynamics (LQCD)~\cite{Wilczek:2002wi}, it is predicted that hadronic matter at extreme high energy density undergoes a phase transition, the QGP~\cite{BraunMunzinger:2008tz}, at a critical (crossover) temperature between 143-171 MeV~\cite{Borsanyi:2010bp, Bernard:2004je, Cheng:2006qk, Bhattacharya:2014ara, Bazavov:2011nk,Ayala:2014jla} at zero baryon chemical potential. The mechanism is driven by the color screening meaning that the screening of the color potential does not allow quarks to be bound into hadrons. This state of matter was able to exist at the quark epoch of the early Universe, around $10^{-12}$ to $10^{-6}$ seconds after the Big-Bang, and perhaps can exist in the core of neutron stars. 

Ultrarelativistic heavy-ion collisions, like those produced at the Relativistic Heavy Ion Collider (RHIC) at Brookhaven National Laboratory and the Large Hadron Collider (LHC) at the European Organization for Nuclear Research (CERN), allow the study of the QCD phase transitions. In \auau collisions at center of mass energy per nucleon pair \snnt{0.2}, experiments at RHIC claimed the discovery of a QGP which behaved as a perfect fluid, and not as the expected gas~\cite{Adams:2005dq,Back:2004je,Arsene:2004fa,Adcox:2004mh}.  This strongly interacting Quark-Gluon Plasma was characterized by a strong collective flow (how does matter produced in the collisions flow collectively?)  and opacity to jets (how does the QGP respond to an energetic parton passing through it?)~\cite{CasalderreySolana:2011us,Dainese:2004te}. These results were later confirmed and further extended in \snnt{2.76} \pbpb collisions  at the LHC. Although several important measurements have been carried out by the LHC high energy physics experiments (CMS, ATLAS and LHCb); the present review focuses on results from ALICE~\cite{0954-3899-32-10-001}, the heavy-ion dedicated experiment at the LHC. 

In this review, we also address the discovery of QGP-like features in small systems which are those formed in high multiplicity \pp and \ppb collisions. This finding itself breaks a paradigm, since prior to the LHC era the measurements in small systems were assumed to represent the vacuum, and therefore, their comparisons with heavy-ion results aimed to extract the genuine QGP features. The progress in the understanding of such effects is widely discussed in Section~\ref{sec:smallsystems}.


\section{Probes of the hot and dense medium created in heavy-ion collisions}
\label{sec:2}

To measure the macroscopic properties of matter that emerge from the fundamental constituents of matter and their interaction within the non-Abelian gauge theory in the regime of extreme energy density, the heavy-ion experimental programs try to identify and quantify the collective phenomena in collisions with the highest energy density. To this end, collisions of high baryonic stable nuclei as \cucu, \auau and \pbpb are chosen. Such collisions create a relative large volume of matter at high energy density allowing the study of the properties that characterize macroscopic quantities of the interacting matter. The large nuclei are accelerated to relativistic speeds at particle accelerators, and to quantify the deviations of these effects, also benchmark measurements are performed in proton-proton (\pp) and proton-nucleus (\pa) collisions, where collective effects were supposed to be absent.

Besides the complex discrimination of the collective nuclear matter effects, some signals of the QGP formation are needed. Such information has to be extracted from the observed hadrons. However, as  hadronization is a non-perturbative QCD phenomenon, the question concerning the posible memory loss of QGP remains open. Nonetheless, we assume that the observed hadrons can give us information of the QGP.

Namely, radial and anisotropic flow (originated from the partonic stage) have been observed in heavy-ion collisions experiments at different $\sqrt{s_{\rm NN}}$. The anisotropy in the spatial distribution of the nucleons participating in the collision is converted into a momentum anisotropy, if sufficient interactions within the medium occur. Hence the azimuthal distribution of the particles in the final state reflects the initial anisotropy of the collision zone and can be used to quantify medium properties. The azimuthal anisotropy of produced particles is commonly characterized by the second Fourier coefficient $v_{2} = \langle \cos[2(\varphi - \Psi_{2})] \rangle$,
where $\varphi$ is the azimuthal angle of the particle momentum, and  $\Psi_{2}$ is the azimuthal angle of the initial-state symmetry plane for the second harmonic \cite{Ollitrault:1992bk}. Nowadays, different methods to measure $v_2$ are employed. Besides  the traditional studies of correlations of particles relative to the reconstructed event plane, two- and multi-particle correlations and cumulants are used in order to disentangle contributions of flow and non-flow effects. 

Strangeness enhancement was proposed as a signature of the transition to QGP~\cite{Rafalski-Muller:82}.  The threshold energy required to produce a pair of $s\bar s$ quarks is just the mass of two strange quarks. Due to the high temperature involved in the QGP phase, the production of $ s\bar s$ pairs via gluon fusion becomes important. With respect to ``vacuum'', the WA97 experiment reported a pronounced enhancement of the hyperon production at central rapidity in \pbpb collisions at \snnt{0.158}~\cite{Andersen1999401}. The results have been supported by experiments at RHIC.  Recently, the LHC experiments found a similar effect not only in \pbpb collisions, but also in high multiplicity \pp and \ppb collisions.

Jets, heavy flavour hadrons and quarkonia, the so-called hard probes, are considered to be the key probes of the hot and dense QCD medium formed
in high energy heavy-ion collisions. Jets originating from fragmentation of hard partons and hadrons containing a heavy quark ({\it c} and {\it b}) are produced
predominantly  in hard scatterings  during the initial phase of the collision. Therefore they experience the entire evolution of the medium 
created in the collision and can probe medium properties.  One of the key methods used to characterise the medium 
(medium density, temperature and transport coefficient) is the study of energy loss of partons traversing the medium. 
In QCD, radiative in-medium energy loss is one of the main contributing mechanisms which depends on the mass and the color 
charge of the given parton. The radiation is suppressed at small angles for heavy quarks due to the dead-cone effect~\cite{Dokshitzer:2001zm} 
and is larger for gluons, which have a larger color charge with respect to quarks in general (Casimir coupling factor). 
The amount of suppression of inclusive particle production relative to pp collisions can be quantified with the nuclear modification factor, \raa, defined as 

\begin{equation}\label{eq:res:1}
R_{\rm AA}(\pt) = \frac{1}{\langle N_{\rm coll} \rangle} \frac{{\rm d}\sigma^{AA \rightarrow a X}/{\rm d}\pt}{{\rm d}\sigma^{pp \rightarrow a X}/{\rm d}\pt}
\end{equation}
where $ a$ denotes the heavy flavour hadrons or quarkonia, $\sigma^{\rm AA}$ and $\sigma^{\pp}$ are the cross section in
\pp and $\rm AA$  collisions, respectively; and  $\langle N_{\rm coll} \rangle$ is the average number of binary
nucleon-nucleon collisions. Studying identified particle production, a hierarchy in the respective nuclear modification factor is expected to be observed when comparing the mostly gluon-originated light-flavor hadrons (e.g. pions) to D and to B mesons. The measurement and comparison of these different probes of the medium provides a unique test of the colour charge and mass dependence of parton energy loss~\cite{Armesto:2003jh}.

Quarkonium states are expected to be suppressed in the QGP, due to colour screening of the force binding the {\it c$\bar{c}$} (or {\it  b$\bar{b}$}) pairs. The measurement of J/$\psi$  production in heavy-ion collisions was therefore proposed as a probe to study the onset of de-confinement already in 1986~\cite{Matsui:1986dk}.  In \pbpb collisions at LHC energy, it is expected that the abundant production 
of charm  quarks in the initial state leads to additional charmonium generation from (re-)combination of {\it c} and {\it $\bar{c}$} quarks along the collision history and/or at hadronisation~\cite{BraunMunzinger:2000px,Schroedter:2000ek} resulting in an enhancement of the J/$\psi$  yield.  In order to measure the effects related to cold nuclear matter (CNM), \ppb collisions have also been studied at the LHC. The study of these effects is important to disentangle hot (QGP related) and cold nuclear matter effects in \pbpb collisions. 
In contrast to J/$\psi$,  bottomonia are considered a cleaner QGP probe as they are less affected by these effects.

Further insights into the medium properties can be obtained by the measurement of azimuthal anisotropy of hard probes in non-central heavy-ion collisions. At low $p_{\rm T}$, the $v_{2}$ of heavy-flavour hadrons  is sensitive to the degree of thermalization of charm and beauty quarks in the deconfined medium. At higher \pt, the measurement of $v_{2}$ carries  
information on the path length dependence of in-medium parton  energy loss. The measurement of heavy-flavour $v_{2}$ offers a unique
 opportunity to test whether also quarks with large mass participate in the collective expansion dynamics and possibly thermalize in the QGP.
 
Electromagnetic probes provided by large \pt particles that do not interact strongly with the medium (like real and virtual photons) constitute control probes for checking our understanding of perturbative QCD in nuclear collisions. On the other hand using low \pt electromagnetic probes,  an indication of the chiral symmetry restoration has been seen in the invariant mass distribution of prompt low mass lepton pairs by means of an enhancement in the continuum yield just below the $\rho/\omega$ resonance~\cite{Schukraft:2013wba}. This was qualitatively in agreement with signals expected from chiral symmetry restoration, where the mass and/or width of hadrons are modified in the vicinity of the QCD phase transition (in this case the $\rho$, observed inside the medium via its two lepton decay).


\section{Soft probes}

To learn about the early state of the system, low \pt ($<2.2$\,GeV/$c$) direct photons are studied. A temperature $T$~=~297$\pm$12$^{\rm stat}$$\pm$41$^{\rm syst}$ MeV has been measured for the 0-20\% \pbpb collisions at $\sqrt{s_{\rm NN}}=$2.76\,TeV~\cite{Adam:2015lda}. Hence, the system at the LHC is hotter than that produced at RHIC, where an early temperature of 239$\pm$25$^{\rm stat}$$\pm$7$^{\rm syst}$ MeV was measured for the same centrality class in \auau collisions~\cite{Adare:2014fwh}. It is worth noticing that such temperatures are already above the one predicted to achieve the QCD phase transition~\cite{Ayala:2014jla}.  The system formed at the LHC is also denser, the average multiplicity per number of participant is twice that measured at RHIC~\cite{Aamodt:2010cz}.

The system expands and cools down, when the inelastic interactions cease the yields of particles are fixed. This is the stage of the so-called chemical freeze-out which is studied using the yields of identified hadrons.  Within 20\% particle ratios, e.g., the proton yield normalized to that of pions, are described by thermal models with a common chemical freeze-out temperature of $T_{\rm ch}\approx$ 156 MeV~\cite{Floris:2014pta}. However, larger deviations are observed for protons and $\rm K^{*0}$, for the latter, this is not a surprise since its mean lifetime is smaller than that of the fireball ($\approx$10 fm/$c$)~\cite{Aamodt:2011mr}, and therefore the resonance yield may deviate from the expected values  due to hadronic processes like re-scattering and regeneration~\cite{Abelev:2014uua}. 

\begin{figure}[t!]
\begin{center}
  \includegraphics[height=.22\textheight]{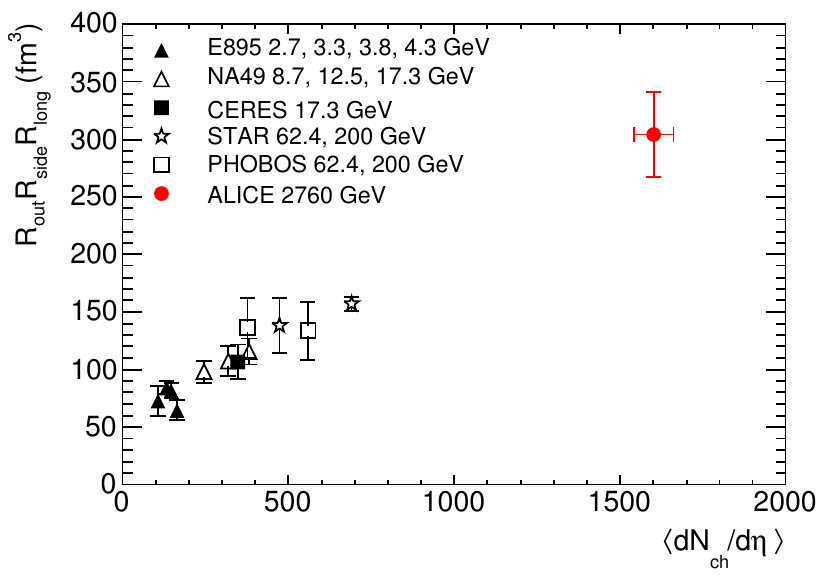}
  \includegraphics[height=.22\textheight]{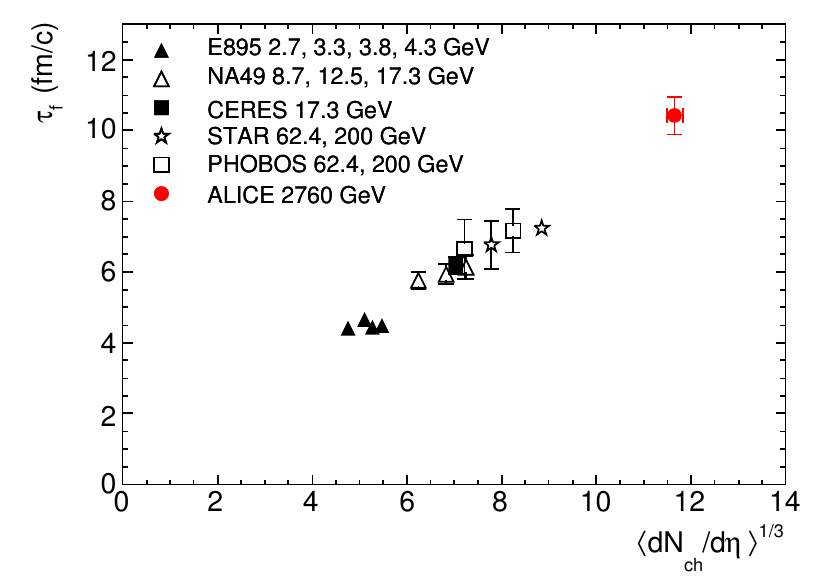}  
  \caption{Product of the three pion HBT radii at $k_{\rm T}=0.3$\,GeV/$c$ (left) and decoupling time (right). The ALICE results (full circles) are compared to those obtained for central Au and Pb collisions at lower energies.}
\label{fig:pbpb:time}
\end{center}
\end{figure}

On the other hand, the measurement of the spatial extent at decoupling time is accesible via intensity interferometry, a technique which exploits the Bose-Einstein enhancement of identical bosons emitted close by in phase space. This approach is known as Hanbury Brown-Twiss analysis (HBT) ~\cite{HanburyBrown:1956bqd}. Such an analysis using identical charged pions has been performed by ALICE. The results give a pion homogeneity volume of $\approx$ 300\,fm$^{3}$ (two times that reported at RHIC) and a decoupling time of $\approx$ 10\,fm/$c$~\cite{Aamodt:2011mr}. The comparisons with results at lower energies are presented in Fig.~\ref{fig:pbpb:time}.

The transverse momentum distributions of identified hadrons contain valuable information about the collective expansion of the system (\pt~$\lesssim$~2\,GeV/$c$), the presence of new hadronization mechanisms like quark recombination (2~$\lesssim$~\pt~$\lesssim$~8\,GeV/$c$)~\cite{Fries:2008hs} and, at larger transverse momenta, the possible modification of the fragmentation due to the medium~\cite{Sapeta:2007ad,Bellwied:2010pr}. ALICE has reported the  transverse momentum spectra of charged pions, kaons and (anti)protons as a function of the collision centrality from low (hundreds of MeV/$c$)~\cite{Abelev:2013vea} to high (20\,GeV/$c$)~\cite{Abelev:2014laa,Adam:2015kca} \pt. 

\begin{figure*}[t!]
\begin{center}
\includegraphics[keepaspectratio, width=0.5\columnwidth]{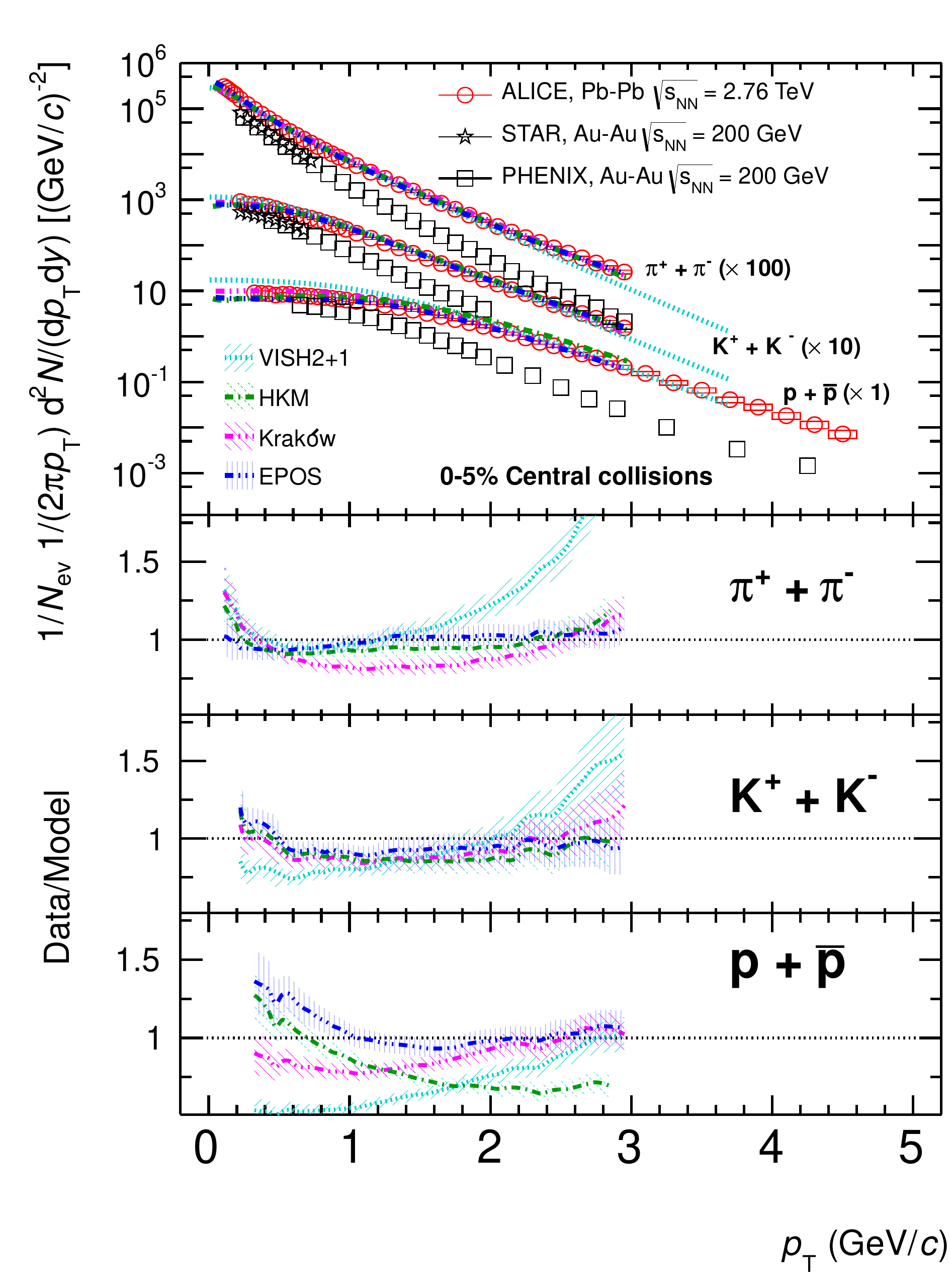}
\includegraphics[keepaspectratio, width=0.5\columnwidth]{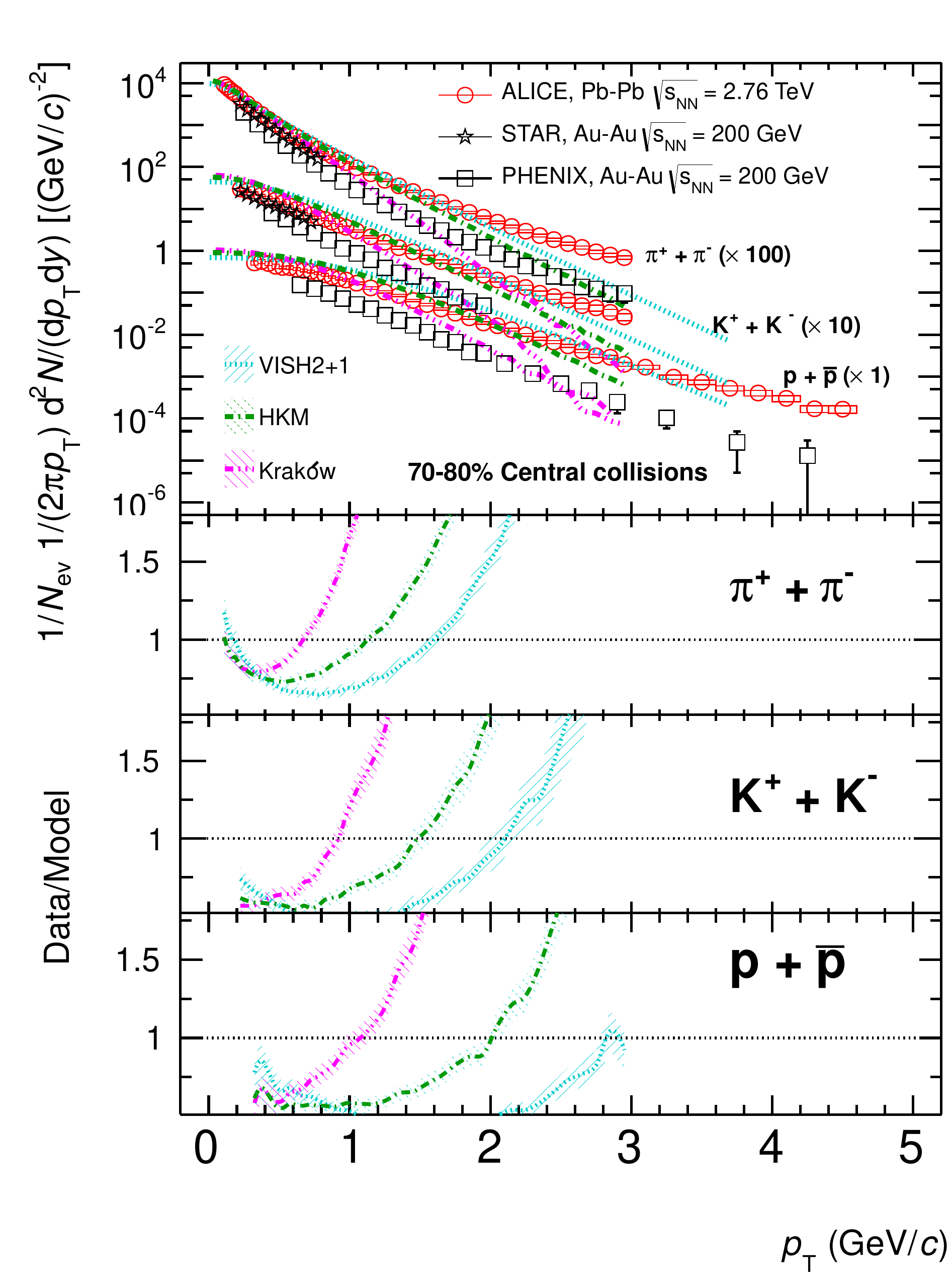}
\caption{\label{fig:pbpb:1} Transverse momentum spectra of charged pions, kaons, and (anti)protons measured in central (left) and peripheral (right) \pbpb collisions at $\sqrt{s_{\rm NN}}=2.76$\,TeV. The systematic and statistical uncertainties are plotted as color boxes and vertical error bars, respectively.}
\end{center}
\end{figure*}

Figure~\ref{fig:pbpb:1} shows that for central \pbpb collisions the low momentum spectra ($<$~2\,GeV/$c$) are well described by hydrodynamic models (within 20\%), except the low \pt ($<$~1\,GeV/$c$) proton yield~\cite{Bozek:2012qs,Karpenko:2012yf,Werner:2012xh,Shen:2011eg}. Models which best describe the data include hadronic rescattering with non-negligible antibaryon annihilation~\cite{Werner:2012xh,Shen:2011eg}. The description of the results by hydrodynamic models is only observed in 0-40\% \pbpb collisions, results for more peripheral collisions disagree with such prediction. This behavior has been recently studied for the average \pt in different colliding systems~\cite{Ortiz:2015cma}. In order to quantify the freeze-out parameters, a simultaneous fit of  the blast-wave function to the low \pt part of the spectra can be performed.  This model assumes a locally thermalized medium, expanding collectively with a common velocity field and undergoing an instantaneous common freeze-out. 

\begin{figure*}[t!]
\begin{center}
\includegraphics[keepaspectratio, width=0.6\columnwidth]{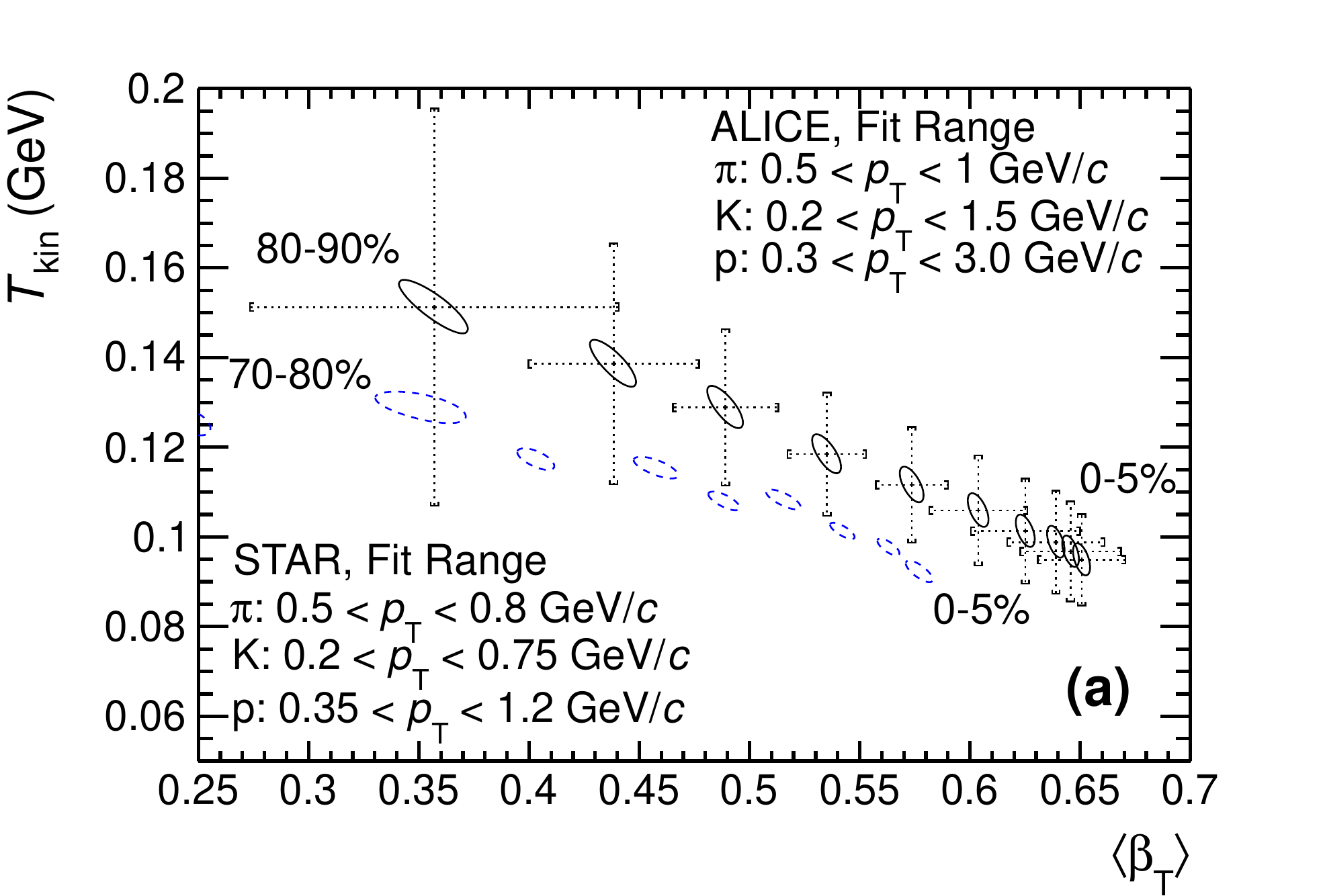}
\caption{\label{fig:pbpb:bw} Results of 
blast-wave fits to pion, kaon and proton spectra in Pb-Pb collisions at 
$\sqrt{s_NN}$=2.76 TeV, compared to similar fits at RHIC energies} 
\end{center}
\end{figure*}

The temperature at the kinetic freeze-out ($T_{\rm kin}$) as a function of the average transverse expansion velocity ($\beta_{\rm T}(r)$) obtained from the blast-wave  analysis~\cite{PhysRevC.48.2462} is shown in Fig.~\ref{fig:pbpb:bw}. At the LHC, the radial flow, $\langle \beta_{\rm T} \rangle$, in the most central collisions is found to be $\approx$ 10\% higher than at RHIC,  while the kinetic freeze-out temperature was found to be comparable to that extracted from data at RHIC, $T_{\rm kin}=$~95\,MeV~\cite{Abelev:2013vea}. From the study of the low \pt particle production we conclude that at the LHC the created system is larger, hotter and longer-lived than that produced at RHIC.

\begin{figure*}[t!]
\begin{center}
\includegraphics[keepaspectratio, width=0.9\columnwidth]{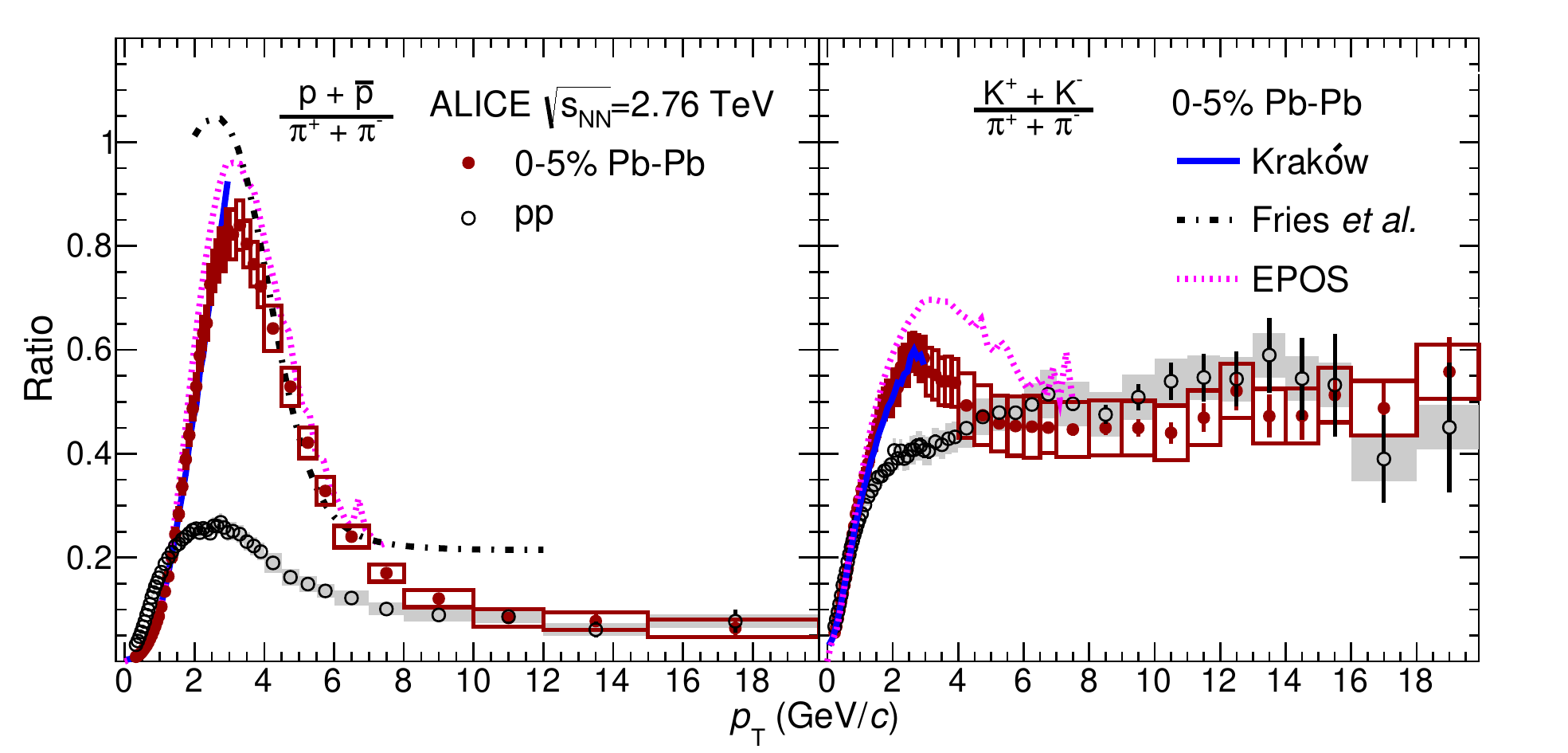}
\caption{\label{fig:pbpb:2} Particle ratios as a function of \pt measured in pp and the most central, 0-5\%, \pbpb collisions. Statistical and systematic uncertainties are displayed as vertical error bars and boxes, respectively. The theoretical predictions refer to \pbpb collisions.}
\end{center}
\end{figure*}

\begin{figure}[t!]
\begin{center}
  \includegraphics[height=.3\textheight]{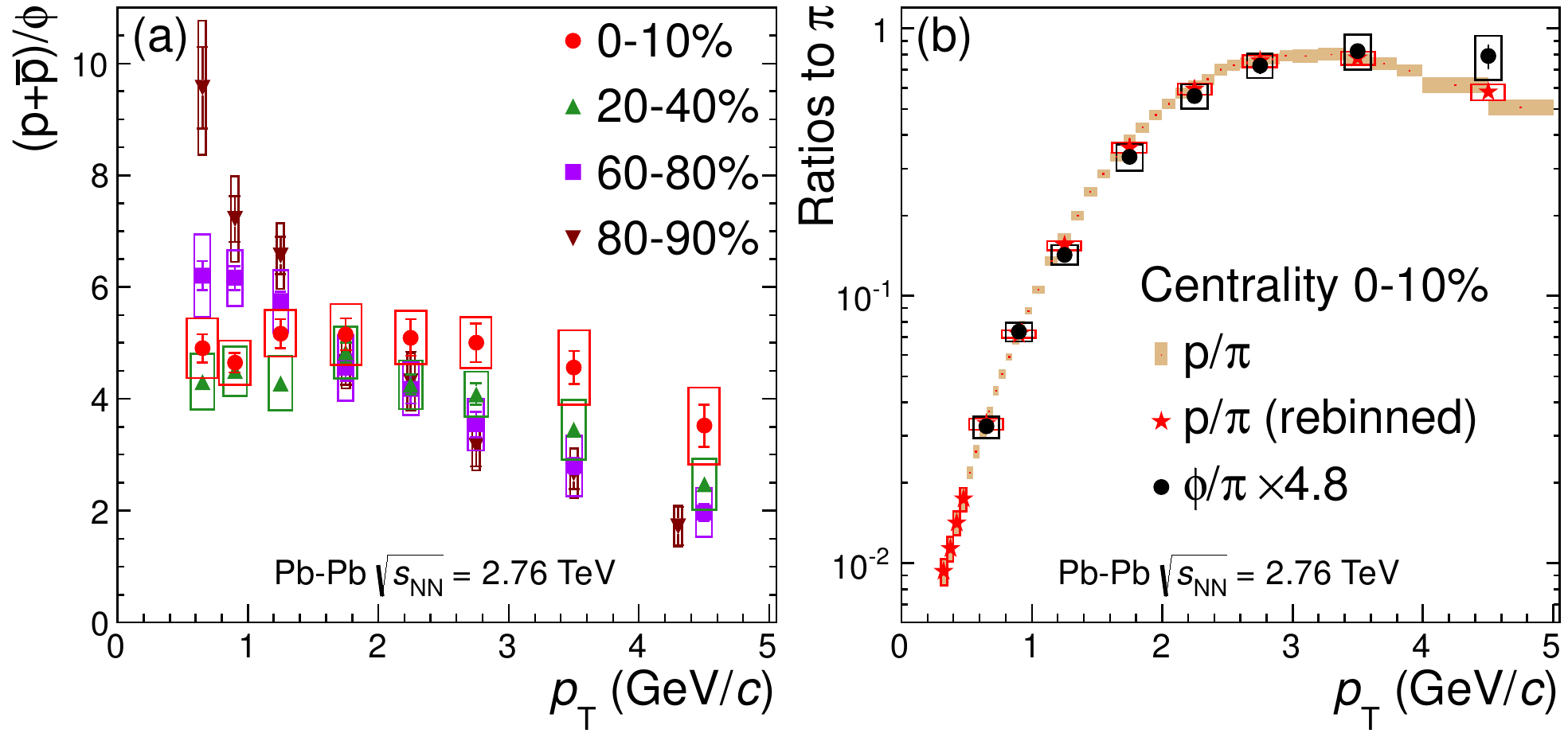}
  \caption{(Color online) Ratio p$/\phi$ as a function of \pt for \pbpb collisions at 2.76 TeV and four centrality classes.  The statistical uncertainties are shown as bars and the total systematic uncertainties (including \pt-uncorrelated and \pt-correlated components) are shown as boxes (left). Ratios of proton and $\phi$-meson yields to charged pion yield as a function of \pt for central \pbpb collisions at 2.76 TeV.  In order to show the shape similarity of the two ratios for $\pt< $3\,GeV/$c$, the $\phi/\pi$ ratio has been scaled so that the $\phi$-meson and proton integrated yields are identical (right).}
\label{fig:pbpb:1b}
\end{center}
\end{figure}

The intermediate \pt is studied with the particle ratios as a function of \pt. Figure~\ref{fig:pbpb:2} shows a comparison between \pp and the most central \pbpb collisions. The proton-to-pion ratio increases from $\approx$ 0.38 to $\approx$ 0.8 going from peripheral (60-80\%) to central (0-5\%) \pbpb collisions at $\pt \approx$ 3\,GeV/$c$, then decreases to the value measured for vacuum fragmentation (\pp collisions) for $\pt>$~10\,GeV/$c$. The result obtained for the most central collisions is similar to that measured at RHIC~\cite{Adare:2013esx,Abelev:2006jr}. The kaon-to-pion ratio also exhibits a bump around $\pt=$~3\,GeV/$c$. This effect is not predicted by quark recombination suggesting that the actual enhancement of the baryon-to-meson ratio is not anomalous and instead it is most likely driven by hydrodynamical flow. Comparisons with theoretical predictions are also shown. The proton-to-pion ratio is only described by quark recombination models for \pt above 3\,GeV/$c$, while hydrodynamical calculations shows a qualitatively good agreement for momentum up to 2\,GeV/$c$. The picture described above is tested by comparing the shapes of the \pt distributions of $\phi$-meson and protons. The results shown in Fig.~\ref{fig:pbpb:1b} indicate that for central \pbpb collisions the shapes of the ratios to pions are the same. The evolution with collision centrality of the $\phi$-meson yield normalized to that of protons as a function of \pt is also shown in Fig.~\ref{fig:pbpb:1b}. For \pt$<$~4\,GeV/$c$ the ratio becomes flat going from the most peripheral to the most central \pbpb collisions. This suggests that the mass, and not the number of quark constituents, determines the spectral shape in central \pbpb collisions. This is in a good agreement with the hydrodynamical interpretation. It has been recently shown that the spectral shapes, studied with the average \pt, exhibit a scaling with the hadron mass (number of constituent quarks) only in the 0-40\% (40-90\%) \pbpb collisions~\cite{Ortiz:2015cma}.

\begin{figure}[t!]
\begin{center}
  \includegraphics[height=.4\textheight]{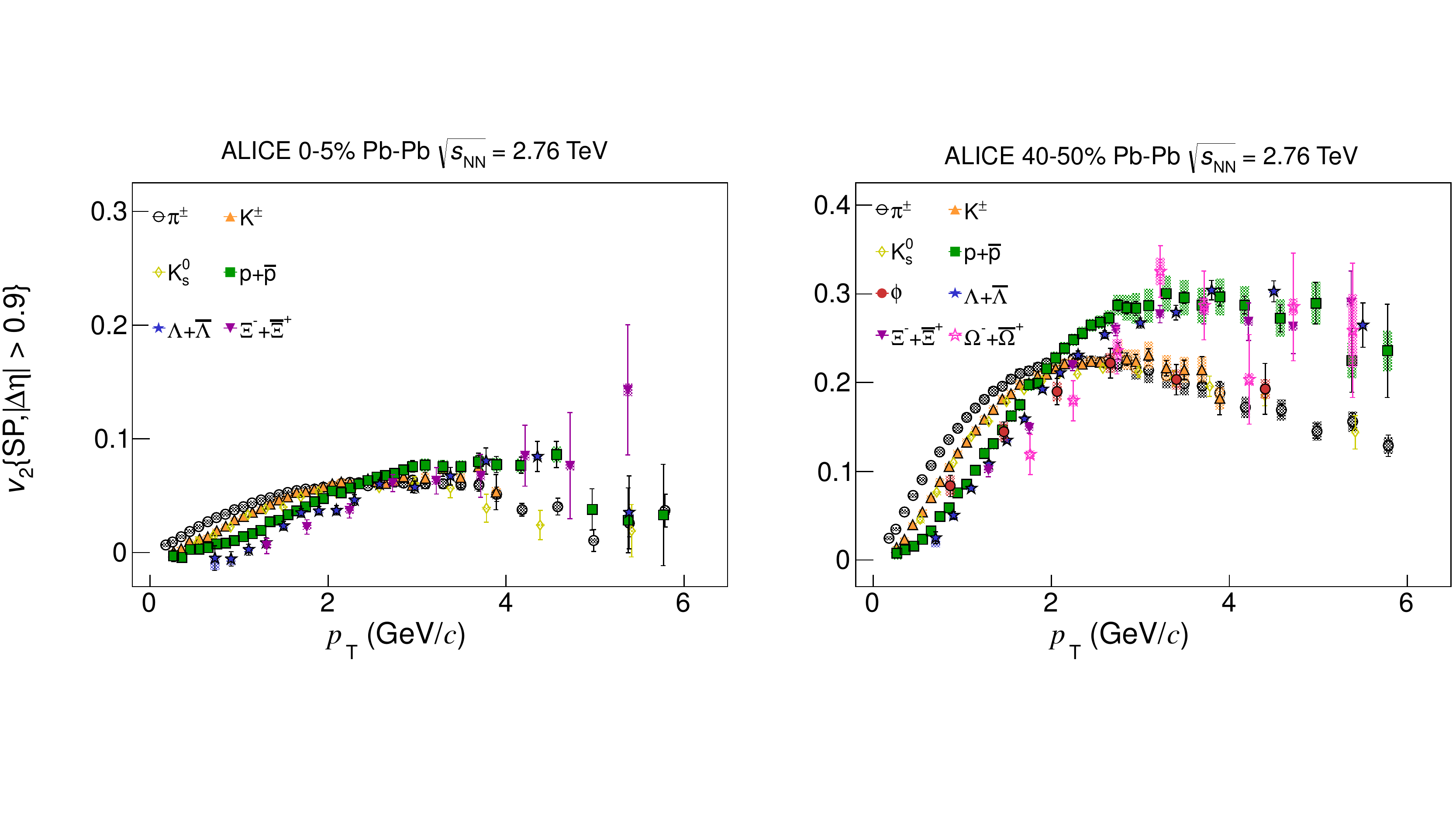}
 \vspace{-2.2cm}
  \caption{(Color online) Elliptic flow coefficient ($v_{2}$) of identified hadrons as a function of \pt measured for central (top) and peripheral (bottom) \pbpb collisions.}
\label{fig:pbpb:1c}
\end{center}
\end{figure}

Using the scalar product method, the elliptic flow for identified hadrons has been  measured~\cite{Abelev:2014pua} over a broad \pt range. Figure~\ref{fig:pbpb:1c} shows $v_{2}$ as a function of \pt for central (0-5\%) and semi-peripheral (40-50\%) \pbpb collisions.  Going from central to semi-peripheral \pbpb collisions $v_{2}$ increases as expected due to the eccentricity increase. For \pt below 2\,GeV/$c$ a mass ordering is observed indicating the interplay between elliptic and radial flow. For higher \pt, the hadron-$v_{2}$ seems to be grouped into baryons and mesons, the exception is the $v_{2}$ of $\phi$-mesons, which for central \pbpb collisions follows that for baryons. This observation indicates that the behavior of $v_{2}$ is driven by the hadron mass and not by the number of quark constituents. ALICE has also reported the violation of the scaling of $v_{2}$ with the number of constituent quarks, such a observation is also against the scenario with quark recombination/coalescence.  

\section{Jet studies}

\begin{figure}[t!]
\begin{tabular}{lr}
\hspace{-1.0cm}
\includegraphics[height=5.5cm]{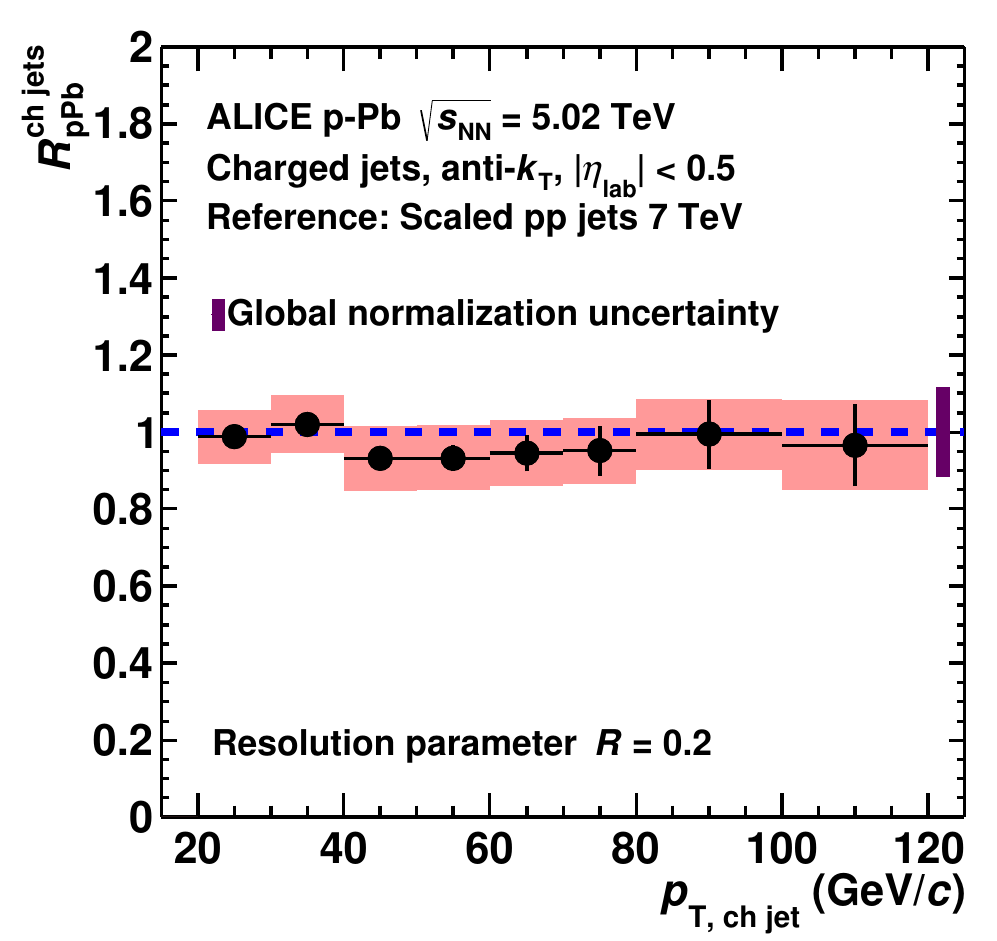}
&
\hspace{-0.5cm}
\includegraphics[height=5.5cm]{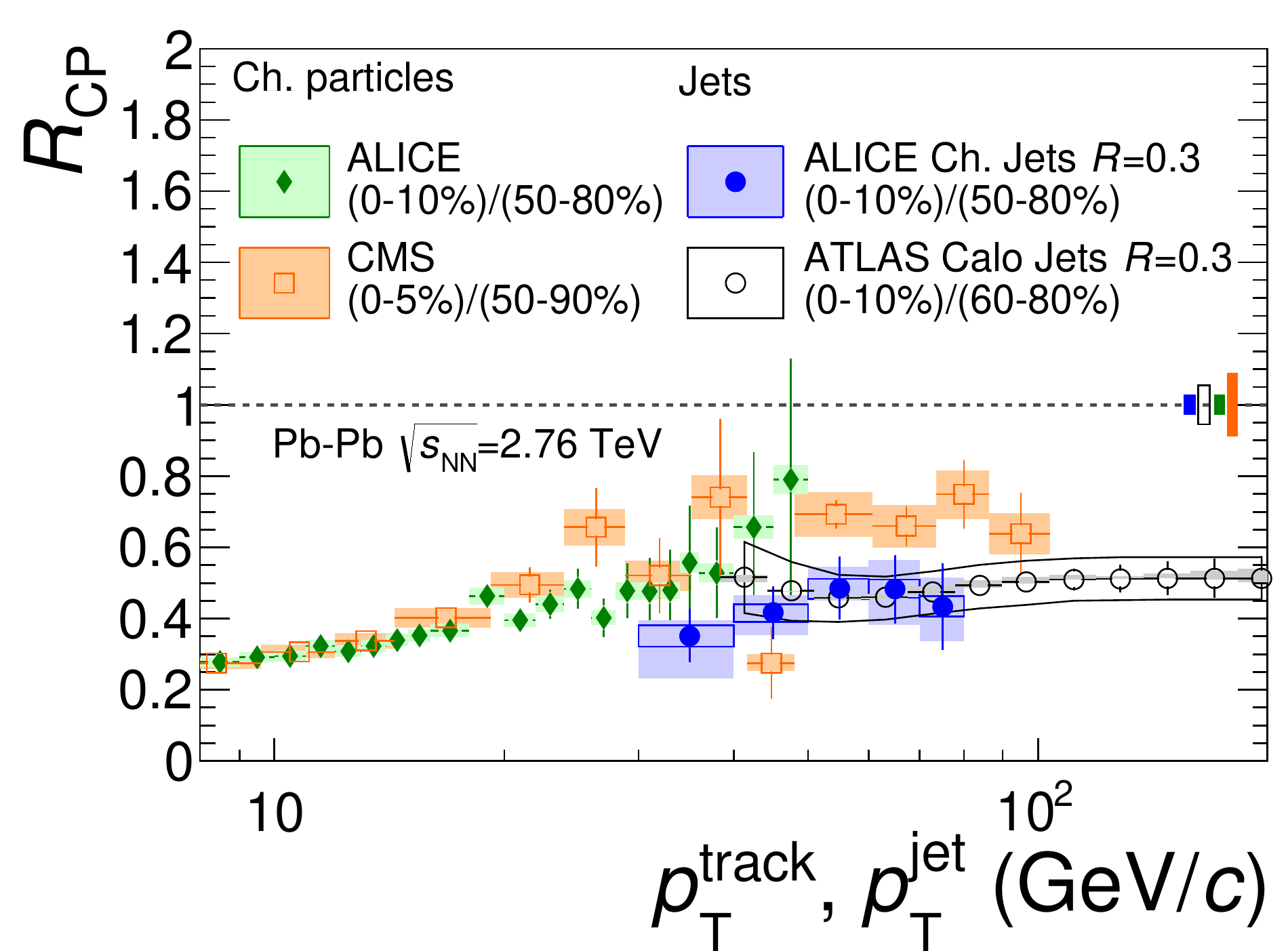}
\end{tabular}
\caption{(Color online). Charged jets \raa for $R$~=~0.2 and \pt leading bias of 5\,GeV/$c$ in  \ppb collisions at $\snnt{5.02}$~\cite{Adam:2015hoa} (left) and in central \pbpb collisions at $\snnt{2.76}$ (right)~\cite{Abelev:2013kqa}. The results from \pbpb collisions are compared with jet \raa measurement from ATLAS~\cite{Aad:2012vca} and charged particle \raa from ALICE~\cite{Abelev:2012hxa} and CMS~\cite{CMS:2012aa}. Note that the underlying parton \pt scale is different for jets and charged particles.}
\label{fig:chargedjetRAA}
\end{figure}

Initial studies of jets and their interaction with QCD matter ({\it ``jet quenching''}) were pioneered at RHIC energy of $\snnt{0.2}$ and were based on measurements of inclusive particle production at large \pt~\cite{Adler:2002xw,Adams:2003kv,Adams:2003im,Abelev:2006db}  and dihadron azimuthal correlations with high-\pt trigger hadrons~\cite{Adler:2002tq, Adams:2005ph} which can be used as a proxy for jets. The jet quenching effect manifested itself in a large  suppression of high-\pt particle production relative to \pp collisions and disappearance  of the away-side correlation peak at intermediate \pT~=~2--6\,GeV/$c$ that is  compensated by increased production of low \pT\ particles. Here we focus on properties of fully reconstructed jets,  which are necessary to complete our understanding of jet quenching process. The jet reconstruction is a challenging task in the environment of heavy-ion collisions due to presence of large and fluctuating underlying background. Pioneering measurements were already performed at RHIC~\cite{Ploskon:2009zd,Bruna:2009em},  however the larger jet cross section at the LHC energies  undoubtedly started a new era. 

The measurements of jet production by ATLAS and CMS in \pbpb collisions showed that 
inclusive jet production is strongly modified relative to \pp collisions~\cite{Aad:2014bxa,Aad:2014wha,Chatrchyan:2014ava}.  
While inclusive jet measurements are sensitive to the average partonic energy loss, dijet measurements probe differences in the quenching 
between two parton showers traversing the medium. The energy imbalance between the leading and sub-leading jet, quantified by a dijet 
asymmetry $A_{\rm{J}}$ parameter~\cite{Aad:2010bu,Chatrchyan:2011sx},  shows a large enhancement in central relative to peripheral \pbpb collisions. In addition, CMS established that the large momentum imbalance is accompanied by a softening of the fragmentation pattern of the sub-leading jet. Consequently, $A_{\rm{J}}$ is recovered when integrating  low \pT\ particles distributed over large angles 
relative to the direction of the sub-leading jet. Detailed investigation of the angular radiation pattern in terms  of  multiplicity, angular and \pT\ spectra 
of the radiation balancing large $A_{\rm{J}}$ was consequently measured by CMS~\cite{CMS:2014uca}.  Differential measurements of correlated jet pair production quantified 
by the rate of neighbouring jets that accompany a jet within a given range of angular distance (\RdR) could be used to quantify fluctuations in the jet energy loss  and help to  discriminate among theoretical  models.  The nuclear modification factor of \RdR\ exhibits a suppression of 0.5--0.7 in central \pbpb collisions, which decreases with increasing nearby jet energy, although this measurement is currently of limited statistical precisions~\cite{Aad:2015bsa}. Although ALICE is not optimized for jet studies, its excellent charged particle tracking from low to high momenta and particle identification capabilities, allow to complement and further extend the measurements from ATLAS and CMS performed for high energetic jets.

\begin{figure}[t!]
\begin{tabular}{lr}
\includegraphics[height=6.0cm]{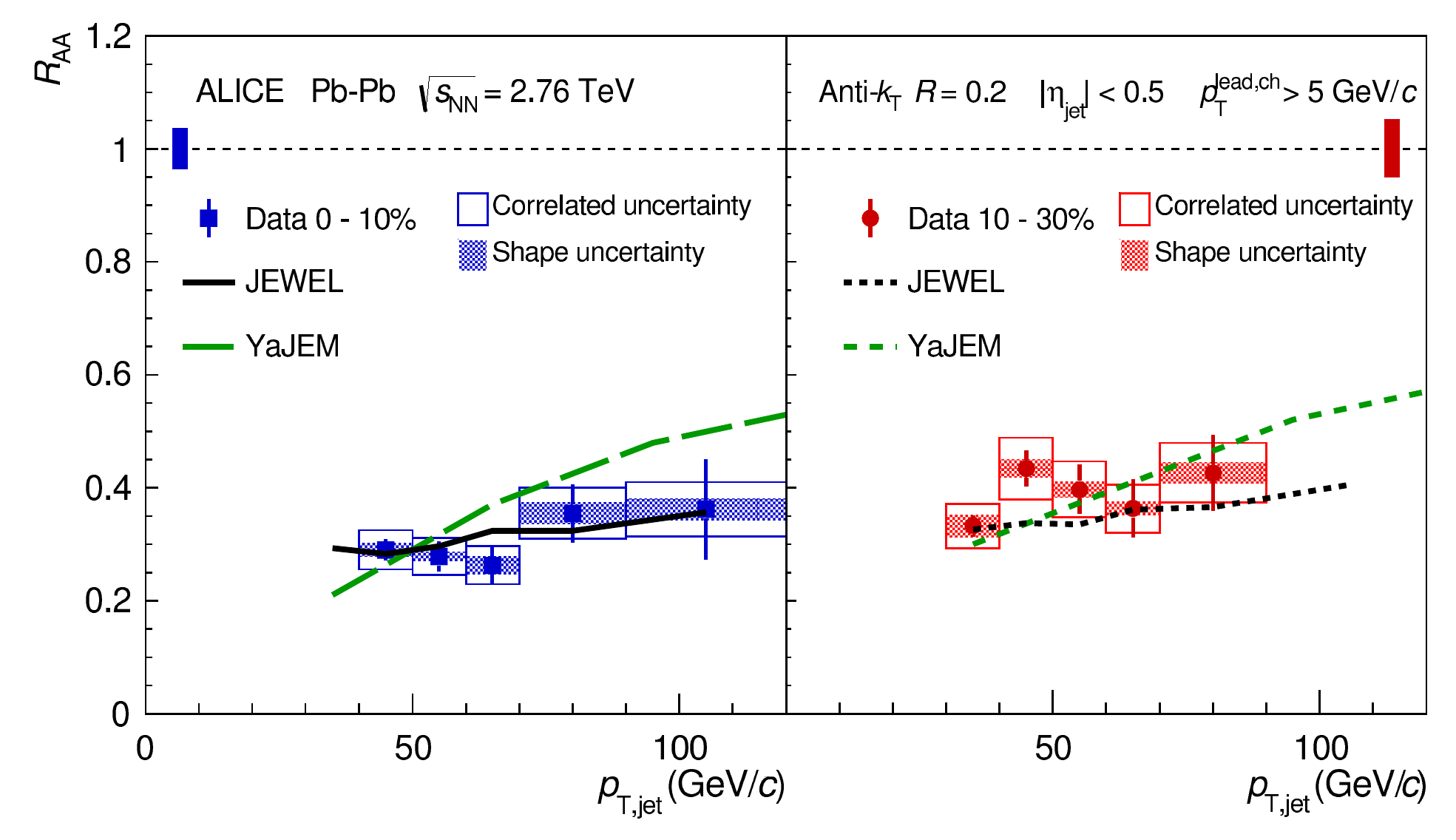}
\end{tabular}
\caption{(Color online). Nuclear modification factor of full jets with $R$~=~0.2 and \pt leading bias of 5\,GeV/$c$ in central \pbpb collisions~\cite{Adam:2015ewa}.}
\label{fig:fulljetRAA}
\end{figure}

To quantify the size of cold nuclear matter effects  on jet production, detailed measurements of jet properties in \ppb collisions at $\snnt{5.02}$ were performed by ALICE. The nuclear modification factor $R_{\mathrm{pPb}}$  for charged jets in Fig.~\ref{fig:chargedjetRAA} (right),  is found to be consistent with unity and points to negligible cold nuclear matter effects in the measured \pt range from 20-120\,GeV/$c$. The measurement as a function of the event centrality, based on the forward neutron energy,~\cite{Adam:2016jfp} corroborates these findings and no statistically significant cold nuclear matter effects are observed even in most ``central'' \ppb collisions.  We remark that due to unavailability of the measured pp reference at $\sppt{5.02}$ in Run I at the LHC, the pp reference used in this measurement is a scaled pp reference from  $\sppt{7}$~\cite{ALICE:2014dla}. ALICE also performed studies of dijet acoplanarity in \ppb collisions~\cite{Adam:2015xea} and also there no significant indications of cold nuclear matter effects are observed and data are found to be consistent with PYTHIA simulations. We can therefore conclude that contributions from initial state effects on studies of jet properties in heavy-ion collisions are small and move to the discussion of jet properties in \pbpb collisions.

Nuclear modification factors of charged  and full jets with $R$~=~0.2--0.3 in central \pbpb collisions at $\snnt{2.76}$ are depicted Fig.~\ref{fig:chargedjetRAA} (right) and in Fig.~\ref{fig:fulljetRAA}, respectively. The data manifest a strong suppression of jet production relative to peripheral \pbpb collisions used as a reference for the charged jet study and relative to the pp reference in case of full jet studies~\cite{Abelev:2013fn}.  Comparing the nuclear modification factor of charged jets to that of charged particles at large \pt one can see that the amount of the suppression is similar although the underlying parton \pt scale is different for inclusive particles and jets. This observation, although it may seem at first glance counterintuitive as one would  expect jets to recover most of the radiated energy, could be explained by the radiation of soft particles at large angles away from the jet axis. The nuclear modification factor of full jets is in Fig.~\ref{fig:fulljetRAA} confronted with two models of jet quenching. The JEWEL model~\cite{Zapp:2012ak} incorporated a microscopical description of the transport coefficient $\hat{q}$ and to determine the initial geometry it uses a combination of Glauber approach and PYTHIA including one dimensional Bjorken expansion of the medium. YAJEM~\cite{Renk:2013rla} is based on 2+1D hydrodynamical description including Glauber Monte-Carlo approach for initial state and leading order QCD calculation to determine outgoing partons.  Both predictions eventually use the Lund hadronization model implemented in PYTHIA. Although both models are conceptually different and in YaJEM the jet quenching effect is a bit stronger with \pt than observed in data,  they describe the measured data well. Larger statistics and more differential jet observables are therefore needed to further constrain models.

\begin{figure}[t!]
\begin{tabular}{lr}
\hspace{-0.5cm}
\includegraphics[height=4.5cm]{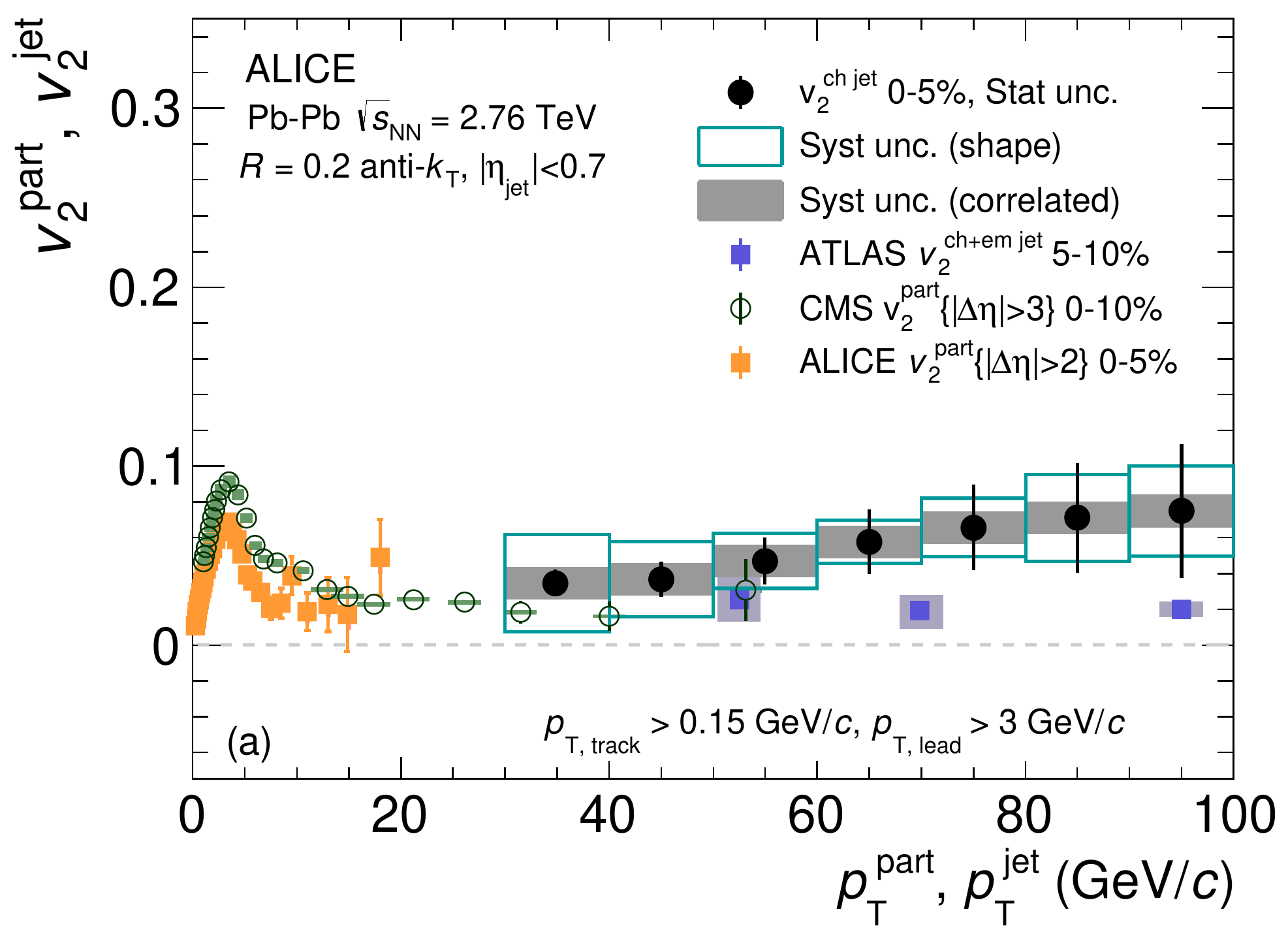}
&
\hspace{-0.5cm}
\includegraphics[height=4.5cm]{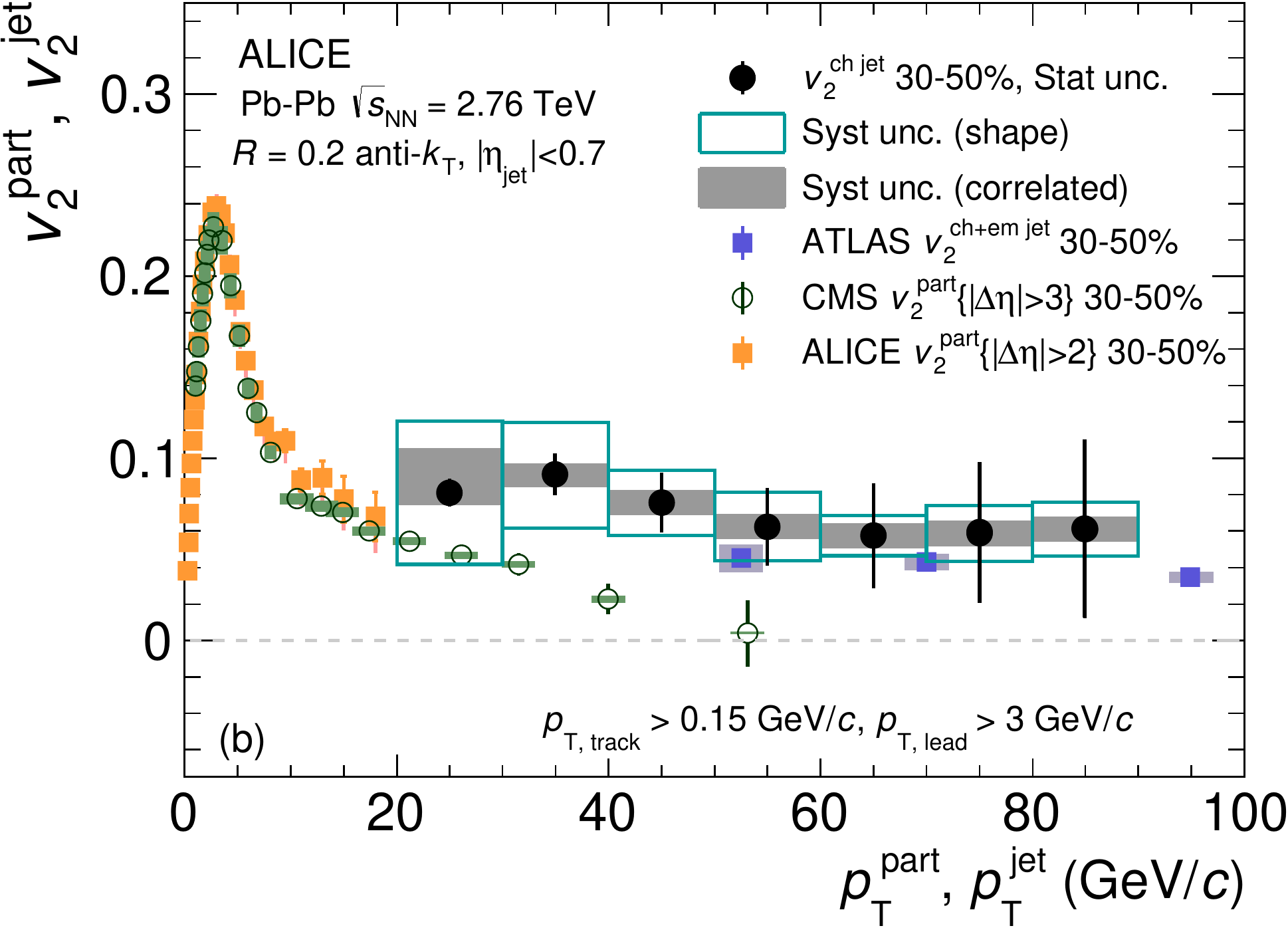}
\end{tabular}
\caption{(Color online). Elliptic anisotropy ($v_2$) of inclusive charged particles and charged jets with the resolution parameter $R$~=~0.2 in central (lef) and semicentral (right) \pbpb\ collisions at $\snnt{2.76}$ measured by ALICE~\cite{Adam:2015mda}. The data are compared with charged particle $v_2$ anisotropies measured by ALICE~\cite{Abelev:2012di} and CMS~\cite{Chatrchyan:2012xq} and calorimetric jets with $R$~=~0.2 measured by ATLAS~\cite{Aad:2013sla}. We remind the reader that the same parton \pT\ corresponds to  different values of single charged particle, charged jet or full jet \pT.} 
\label{fig:v2jetALICE}
\end{figure}

In order to study details of path-length dependence of energy loss, ALICE performed studies of elliptic anisotropy of inclusive charged jets~\cite{Adam:2015mda} and semi-inclusive distributions of recoil jets~\cite{Adam:2015doa} which complement and further extend earlier studies of elliptic anisotropies of inclusive high-\pt  particles and modification of away-side di-hadron correlations~\cite{Aamodt:2011vg}. For collisional energy loss, the path length dependence is expected to be linearly proportional to the length traversed by the parton in medium, while for radiative energy loss, where in addition  interference effects play a role, the dependence can be quadratic. In AdS/CFT class of models  an  even stronger dependence on path length traversed is predicted. In Figure~\ref{fig:v2jetALICE} the measurement of elliptic anisotropy $v_2$ for charged jets with the resolution parameter $R$~=~0.2 is shown in central and semi-central \pbpb collisions. The data show significant positive $v_2$ value in semi-central \pbpb collisions pointing to the path length dependence of jet suppression. In central collisions the current uncertainties on the measurement do not allow to draw a definite conclusion, although the $v_2$ magnitude is also positive. These data are also compared to $v_2$ for full jets measured by ATLAS~\cite{Aad:2013sla} and inclusive charged particles~\cite{Abelev:2012di,Chatrchyan:2012xq}. Although these measurements cannot be directly compared quantitatively due to  different \pt scales and centrality selections, qualitatively they agree and provide a clear evidence of path-length dependent parton energy loss.

\begin{figure}[t!]
\begin{tabular}{lr}
\hspace{-1.0cm}
\includegraphics[height=5.0cm]{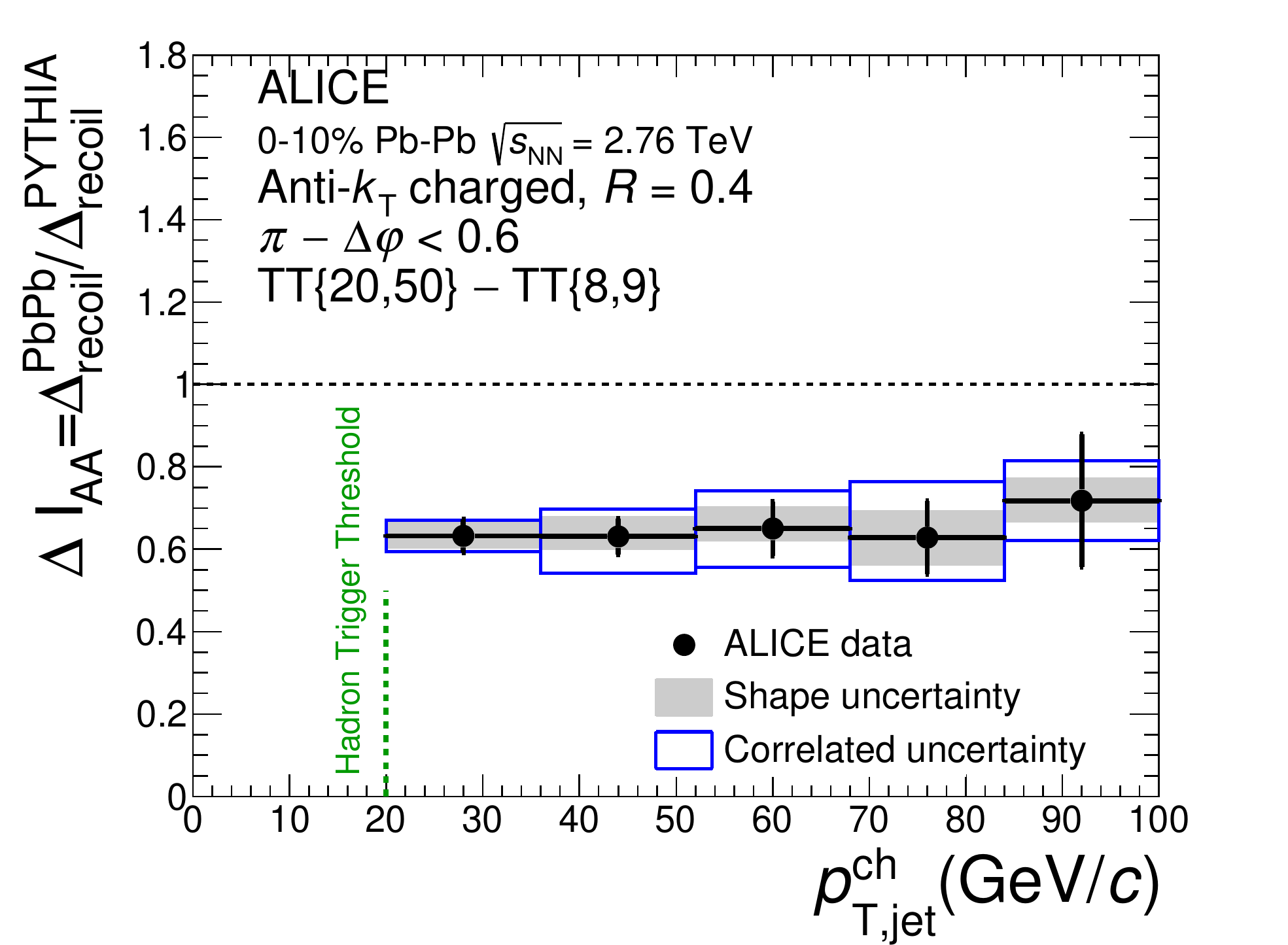}
&
\hspace{-0.5cm}
\begin{minipage}{0.5\textwidth}
\vspace{-9.5cm}
\includegraphics[height=10.0cm]{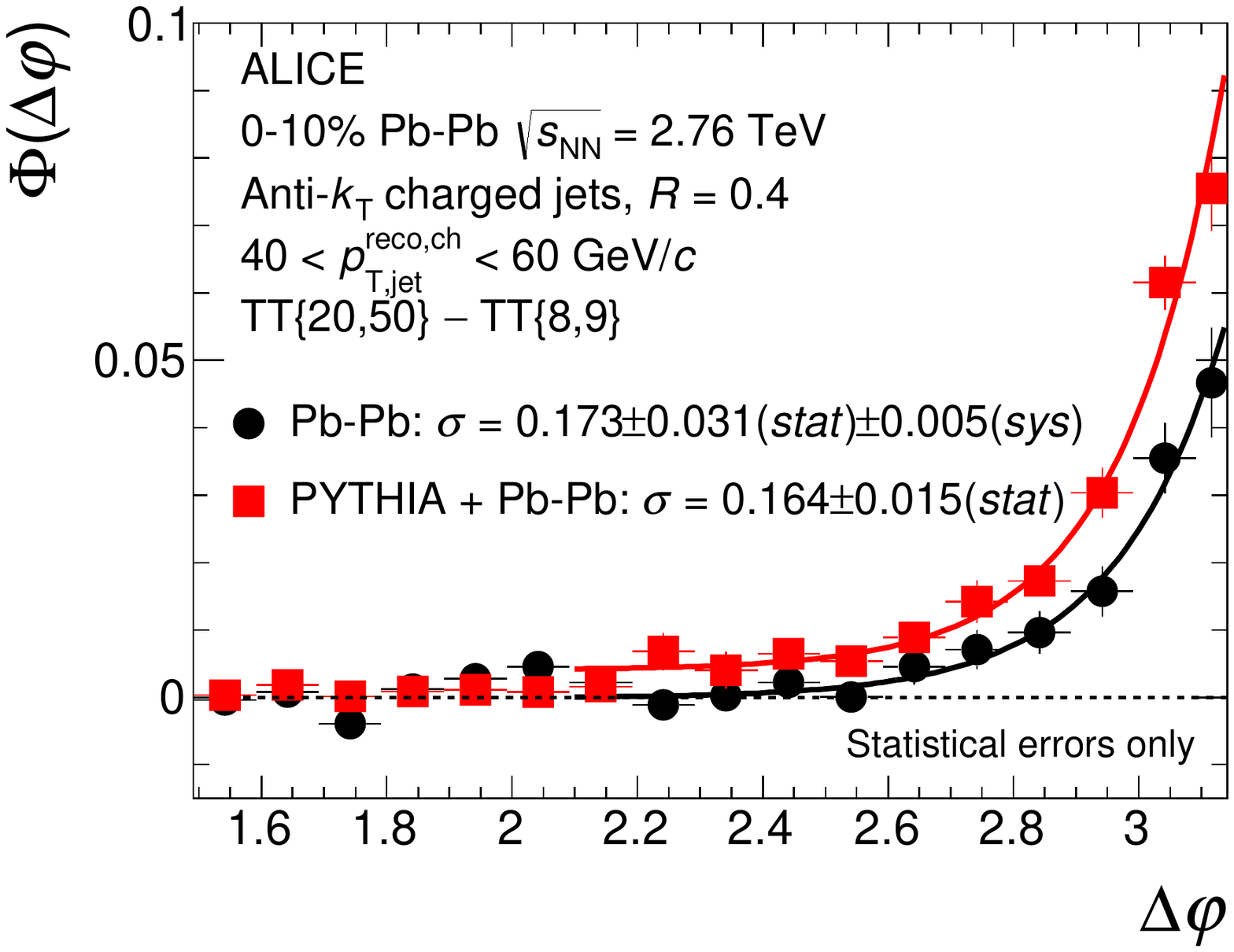}
\end{minipage}
\end{tabular}
\caption{(Color online). Left: the nuclear modification factor $I_{\rm{AA}}$ of recoiling jets in central \pbpb collisions at $\snnt{2.76}$ measured by ALICE~\cite{Adam:2015doa}. Right: azimuthal distribution of recoiling jets relative to trigger hadron orientation in \pbpb (squares) data and PYTHIA (circles) together with a Gaussian fit (lines)~\cite{Adam:2015doa}.}
\label{fig:hjetALICE}
\end{figure}

Measurements of semi-inclusive distributions of jets recoiling from a hard trigger particle in heavy-ion collisions offer a unique approach to jet quenching studies that is calculable in perturbative QCD. This method has been pioneered by the  ALICE experiment~\cite{Adam:2015doa} and recently followed by STAR~\cite{Jacobs:2015srw} although with a technically different implementation of the signal extraction. The essence of this method is to measure semi-inclusive distributions of charged jets recoiling from a charged trigger hadron with a transverse momentum $p_{\rm T}^{\rm trig}$ in a given interval (``TT'') normalized by the number of trigger hadrons (\Ntrig) which can be written as:

\begin{equation}
\frac{1}{\Ntrig}\dNjetdpT\Bigg\vert_{\pTtrig\in{\mathrm{TT}}} = \left(
\frac{1}{\sigma^{\AAtoh}} \cdot
\frac{\rm{d}^2\sigma^{\AAtohjet}}{\mathrm{d}\pTjetch\rm{d}\eta_\mathrm{jet}}\right) 
\Bigg\vert_{\pTh\in{\mathrm{TT}}},
\label{eq:hjet}
\end{equation}
\noindent
where $\sigma^{\AAtoh}$ is the cross section to generate a hadron within the TT trigger interval, $\rm{d}^2\sigma^{\AAtohjet}/\rm{d}\pTjetch\rm{d}\eta$ is the differential cross section for production of a hadron in the TT interval and a recoiling charged jet, and \pTjetch\ and $\eta_\mathrm{jet}$ are the charged jet transverse momentum and pseudorapidity, respectively.  To remove background contributions, ALICE introduced a $\Delta_{\rm recoil}$ observable defined as a difference of semi-inclusive recoil jet distributions from Eq.~(\ref{eq:hjet}) for the signal (\TTSig) and reference trigger particle (\TTRef) classes:
\begin{equation}
\Drecoil = 
\frac{1}{\Ntrig}\dNjetdpT\Bigg\vert_{\pTtrig\in{\TTSig}} - \frac{1}{\Ntrig}\dNjetdpT\Bigg\vert_{\pTtrig\in{\TTRef}},
\label{eq:drecoil}
\end{equation}
with the signal class \TTSig\ having higher \pT\ trigger hadrons than the \TTRef\ class. The \TTSig\ class  corresponds to scattering processes with large $Q^2$ and consequently the associated recoiling jets have a harder \pT\ spectrum than those from  the \TTRef\ class, while the positive part of the jet spectra is in both cases populated by random matching of the trigger hadron and background jets. 
The nuclear modification factor $I_{AA}$ of the recoiling charged jets in \pbpb collisions at $\snnt{2.76}$~ for the resolution parameter $R$~=~0.4 is shown in Fig.~\ref{fig:hjetALICE} (left) 
~\cite{Adam:2015doa}.  
The recoiling jet spectra in \pbpb collisions are divided by the reference \pp spectra from PYTHIA due to the lack of statistics in measured \pp data at the same energy.  Detailed studies in pp collisions at $\sqrt{s}$~=~7~TeV confirmed that PYTHIA is a reliable choice for the \pp reference of recoil charged jet spectra. The data show a significant suppression by up to a factor of two of recoiling jet yield relative to \pp reference for studied range of $R$~=~0.2--0.5~\cite{Adam:2015doa} indicating  that the medium-induced energy loss is radiated at large angles.

To further explore nature of jet quenching, the deflection of the recoil jet axis relative to the signal trigger hadron was studied in~\cite{Adam:2015doa}. The azimuthal distributions of jets  with \pT~=~40--60\,GeV/$c$ displayed in Fig.~\ref{fig:hjetALICE} (right) do not show any significant medium-induced acoplanarity in line with direct photon-jet~\cite{Chatrchyan:2012gt} and dijet correlation studies~\cite{Aad:2010bu}.  Besides the acoplanarity itself it is important to investigate the rate of large angular deviations in tails of the $\Delta\phi$ distribution which is expected to be sensitive to the quasi-particle nature of the medium arising predominantly from single hard (Moli\`ere) scattering. Within the limited Run I statistics the data shows no evidence for Moli\`ere scattering~\cite{Adam:2015doa}.

\begin{figure}[t!]
\begin{center}
\includegraphics[width=6cm, height=6cm]{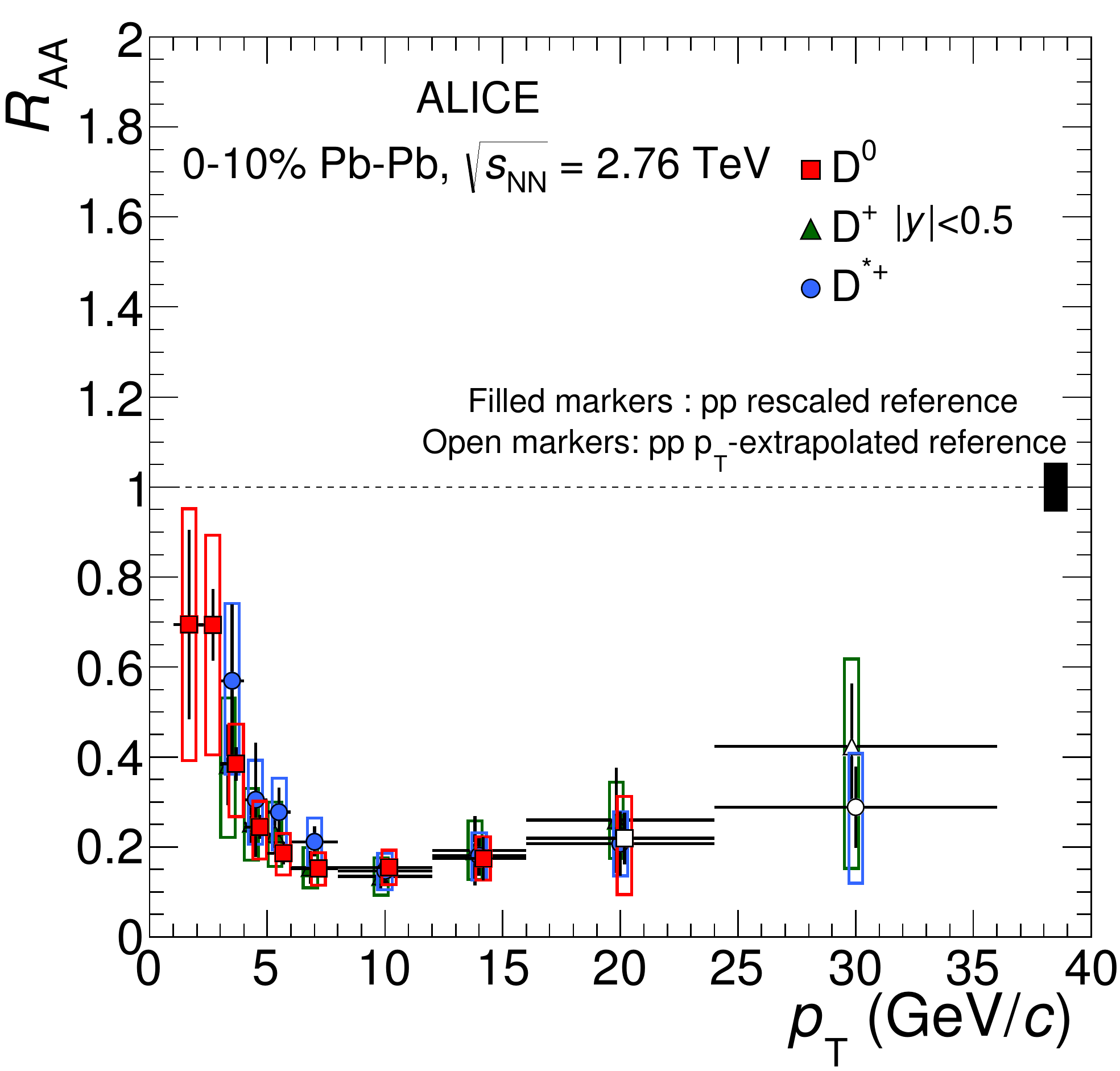}
\includegraphics[width=6cm, height=6cm]{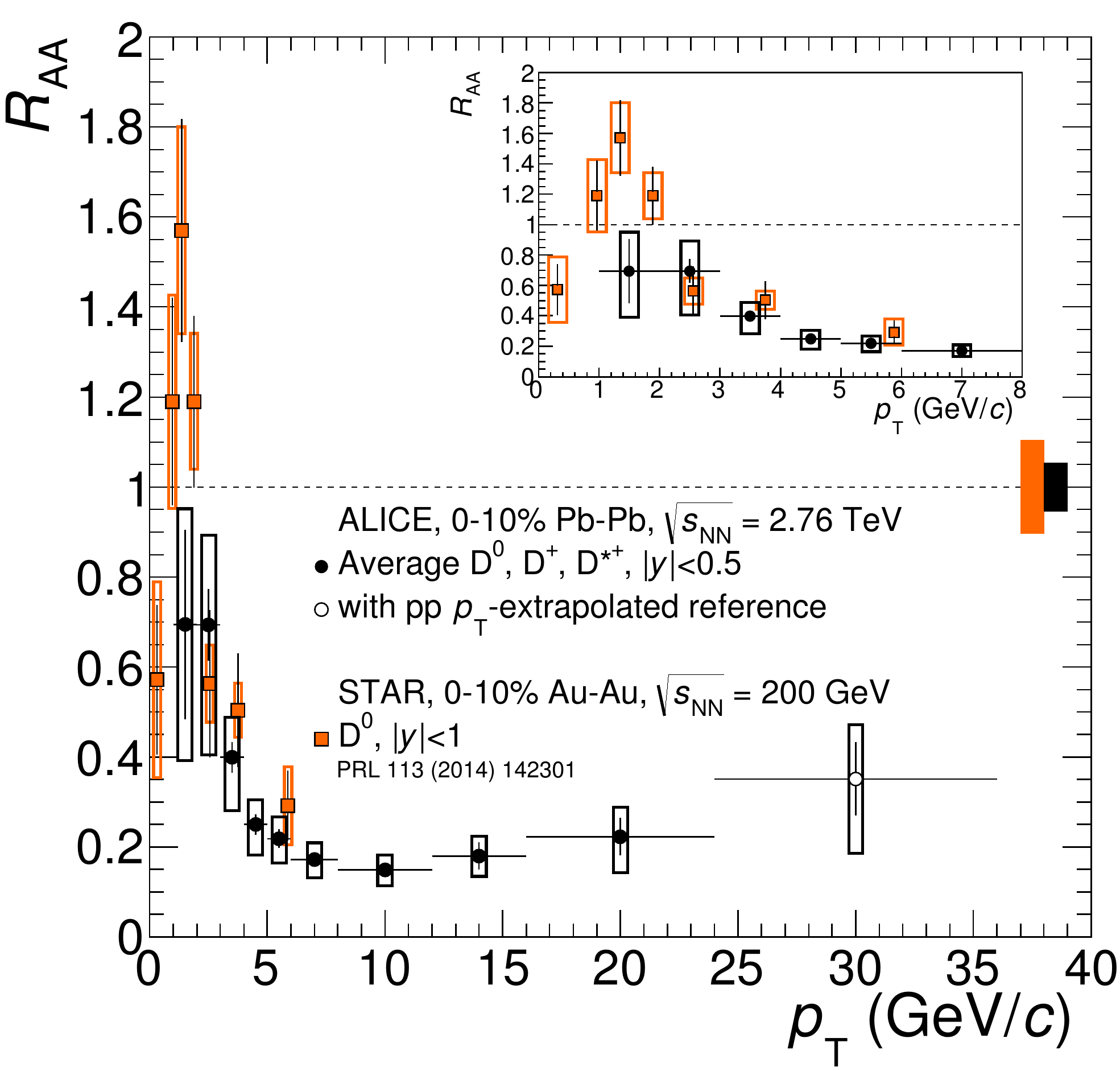}
\caption{(Color online). Left: $R_{\rm AA}$ of  $\rm D^{0}$, $\rm D^{+}$, $\rm
D^{*+}$ and $\rm D^{+}_{s}$ mesons  as a function of $p_{\rm T}$ in
central collision ~\cite{Adam:2015jda}. Right: Average D meson \raa as a function of \pt in 0-10\% centrality compared to D$^{0}$ \raa measured by STAR in \auau collisions at \snnt{0.2}~\cite{Adamczyk:2014uip}.}
\label{fig1:d}
\end{center}
\end{figure}

\section{Open heavy flavour and quarkonia production}

The measurement of open-heavy-flavour production in \pp collisions provides itself important tests of our understanding of various aspects of QCD. However, here we used it as a reference in order to disentangle the genuine properties of the strongly interacting QCD medium created in heavy-ion collisions. 

The nuclear modification factors of D mesons~\cite{Adam:2015sza,Adam:2015jda} and electrons from heavy flavour decays~\cite{Abelev:2012qh} at mid-rapidity and of muons from heavy flavour decays at forward rapidity have been measured with ALICE: all of the results showed a strong reduction of the yields at large transverse momenta (\pt~$\geq$~5\,GeV/$c$) in the most central collisions. Figure~\ref{fig1:d} (left) displays the nuclear modification factor of $\rm D^{0}$, $\rm D^{+}$ and $\rm D^{*+}$ mesons  as a function of $p_{\rm T}$ in the most central collisions~\cite{Adam:2015sza}. The $R_{\rm AA}$ values of the
three D meson species are compatible within uncertainties. A suppression up to a factor five is seen at $p_{\rm T} \sim$~10\,GeV/$c$. Also, the first measurement of
$\rm D^{+}_{s}$  $R_{\rm AA}$ in heavy-ion collisions has been done with ALICE ~\cite{Adam:2015jda}. In the highest measured $p_{\rm T}$ bin (8--12\,GeV/$c$), 
the $R_{\rm AA}$ of $\rm D^{+}_{s}$ mesons is compatible with that of non-strange charmed mesons. At lower $p_{\rm T}$, the $R_{\rm AA}$ of $\rm D^{+}_{s}$ seems to increase, but with the current
statistical and systematic uncertainties no conclusion can be drawn on the expected enhancement of $\rm D^{+}_
{s}$-meson production with respect to that of non-strange D mesons at low \pt, due to
c-quark coalescence with the abundant strange quarks \cite{He:2012df}. In the right panel of Fig.~\ref{fig1:d}, the average D meson $R_{\rm AA}$ in the 0-10$\%$ centrality is compared to the D$^{0}$ $R_{\rm AA}$ measured by  STAR in \auau collisions at $\sppt{0.2}$~\cite{Adamczyk:2014uip}. The $\rm D$ meson $R_{\rm AA}$ measured at two energies are compatible within uncertainties for $\pt$~$>$~2\,GeV/$c$.

\begin{figure}[t!]
\begin{center}
\includegraphics[width=6cm, height=6cm]{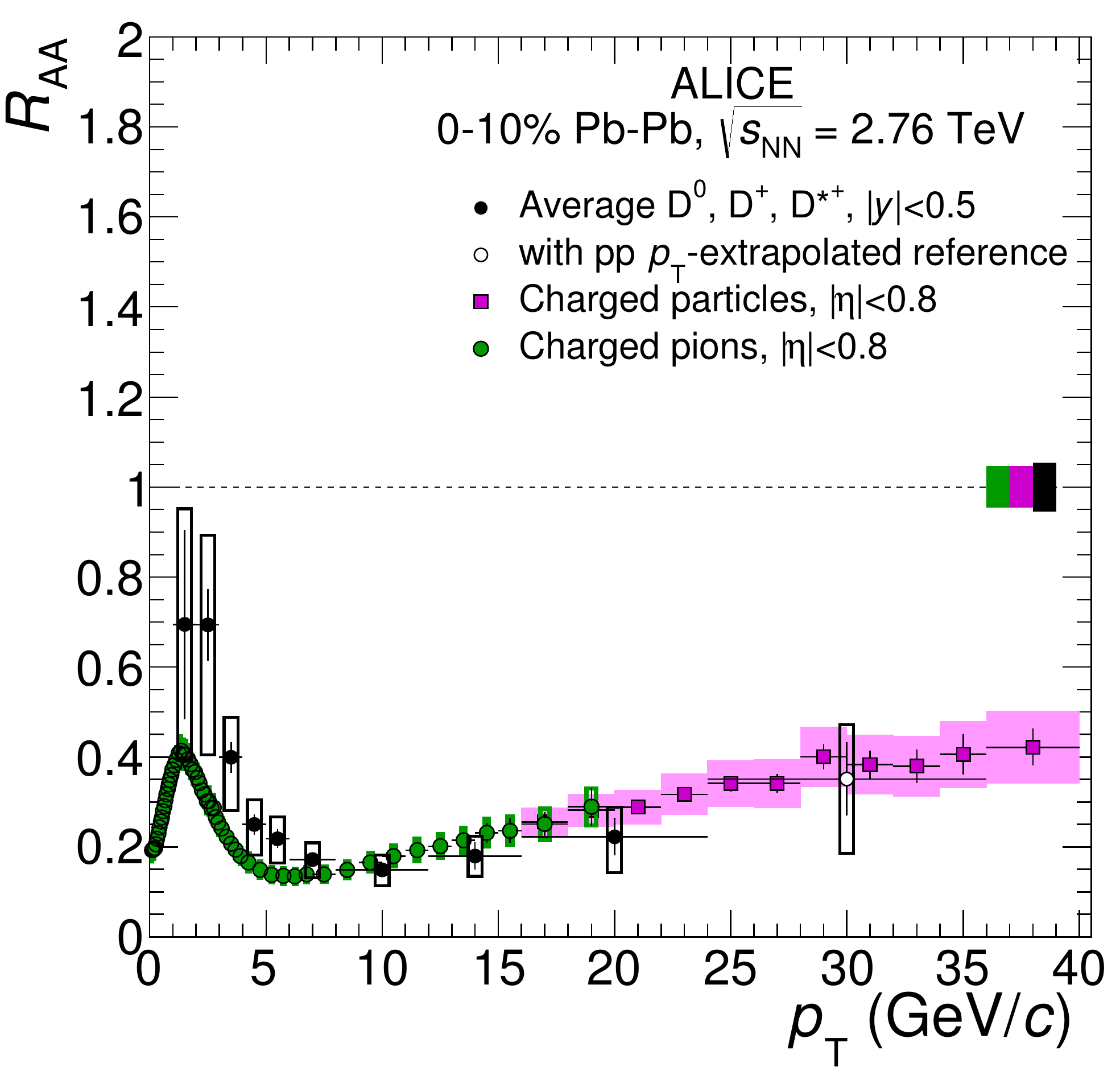}
\includegraphics[width=6cm, height=6cm]{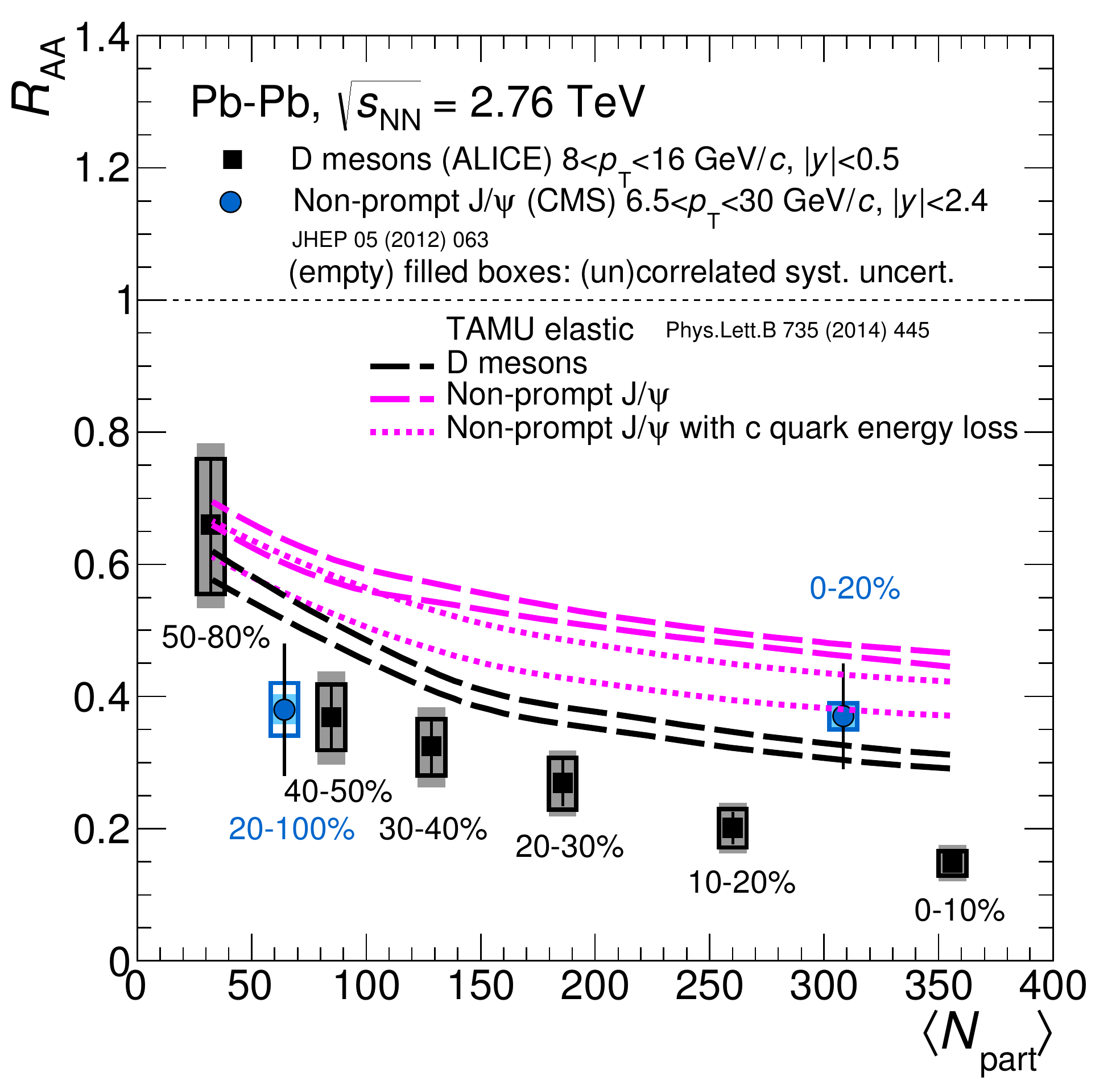}
\caption{(Color online). D meson nuclear modification factor, $R_{\rm AA}$, in \pbpb
  collisions at $\snnt{2.76}$. Left: $R_{\rm AA}$ as a function
of \pt compared to charged hadrons and pions. Right: $R_{\rm AA}$
as a function of $N_{\rm part}$ ~\cite{Adam:2015nna} compared to
non-prompt J/$\psi$ measured by the CMS collaboration~\cite{Chatrchyan:2012np}.}
\label{charged}
\end{center}
\end{figure}

Figure~\ref{charged} (left) shows that, within uncertainties and in the measured \pt interval, the D meson nuclear modification factor is similar to that of charged pions and
inclusive charged particles. It should be noted that the $R_{\rm AA}$ of D mesons and pions is also sensitive
to the shape of the parton momentum distribution and their
fragmentation functions. Model calculations including those effects
and a colour-charge hierarchy in parton energy loss are able to
describe the measurements~\cite{Djordjevic:2013pba}. 
In Fig.~\ref{charged} (right), $R_{\rm AA}^{\rm D}$ as a function of
collision centrality (quantified by the average number of participant
nucleons)~\cite{Adam:2015nna} is shown. This measurement is compared with results from the CMS collaboration of
non-prompt J/$\psi$~\cite{Chatrchyan:2012np} and theoretical predictions~\cite{Andronic:2015wma,Nahrgang:2013xaa}. For D mesons, a smaller suppression in
peripheral than in central collisions is observed. A larger
suppression in central collisions is seen for D mesons than
for non-prompt J/$\psi$, indicating a different energy loss for charm
and beauty quarks. This observation is
supported by predictions from energy loss models, where the difference
between the $R_{\rm AA}$ of D and
B mesons arises from the different masses of $c$ and $b$ quarks. 

\begin{figure}[t!]
\begin{center}
\vspace{-0.5cm}
\includegraphics[width=11cm, height=5.5cm]{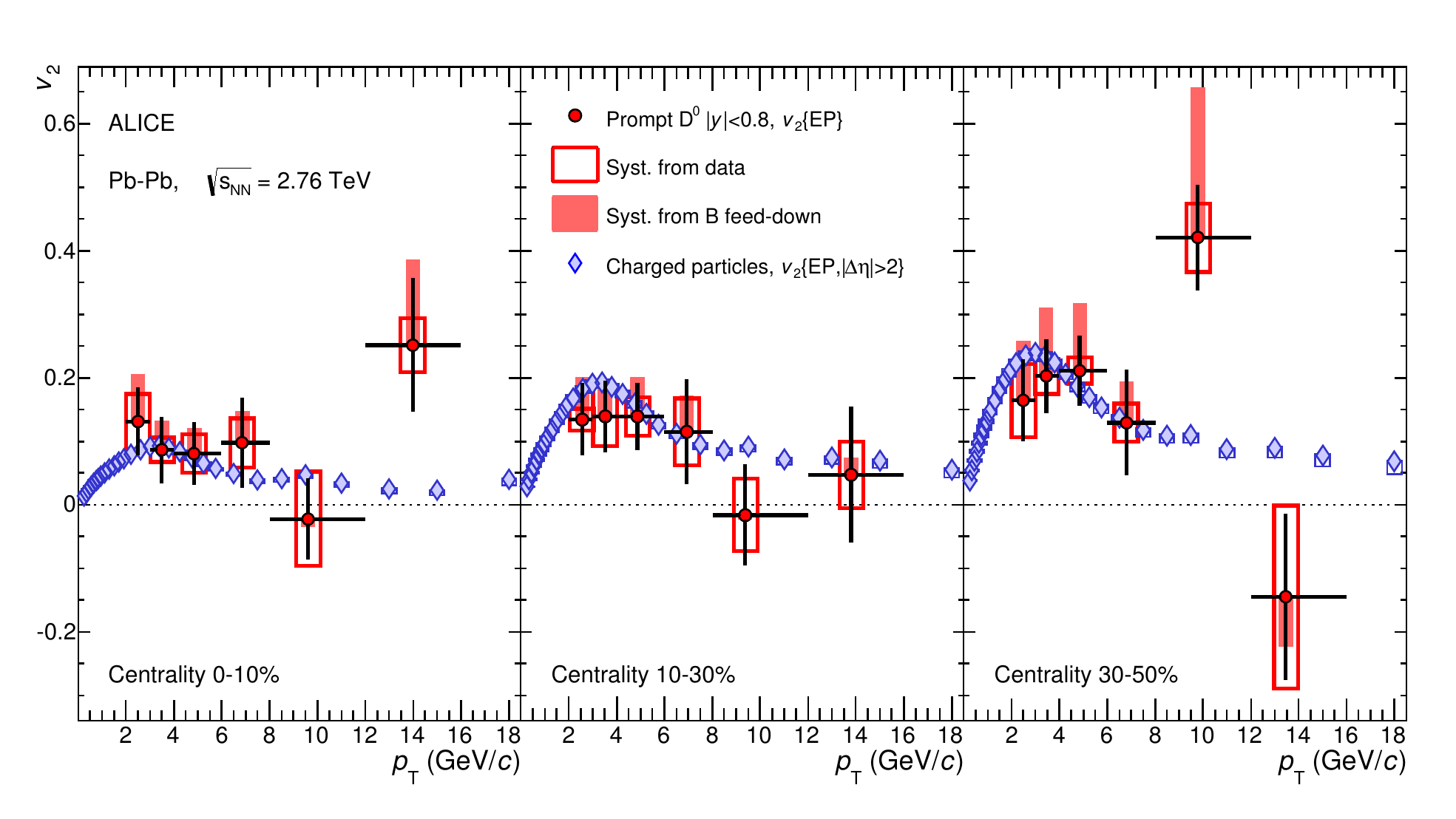}
\caption{(Color online). $\rm D^{0}$ meson $v_{2}$ as a function of \pt in
  three centrality ranges and  compared with the $v_{2}$ of charged
  particles~\cite{Abelev:2013lca,Abelev:2014ipa}.}
\label{v2charged}
\end{center}
\end{figure}

The $v_{2}$ of prompt $\rm D^{0}$, $\rm D^{+}$ and $\rm D^{*+}$ mesons at mid-rapidity was measured in three centrality classes: 0-10\%, 10-30\% and 30-50\%~\cite{Abelev:2013lca,Abelev:2014ipa}. The comparisons of the D meson $v_{2}$ as a function of \pt with  analogous results for inclusive charged particles are shown in Fig.~\ref{v2charged}. The $v_{2}$ decreases from peripheral to central collisions as expected due to the decreasing initial geometrical anisotropy.

The average of the  $v_{2}$ of $\rm D^{0}$, $\rm D^{+}$ and $\rm
D^{*+}$ in the centrality class 30-50$\%$ is larger than zero with
5$\sigma$ significance in the range 2~$<$~\pt~$<$~6\,GeV/$c$. A
positive $v_{2}$ is also observed for $p_{\rm T} >$~6\,GeV/$c$, which
most likely originates from the path length dependence of the
in-medium partonic energy loss, although the present statistics
does not allow to give a firm conclusion on this. The measured D meson
$v_{2}$ is comparable in magnitude with that of the charged particles,
which are mostly light-flavour hadrons. This result indicates that
low-\pt charm quarks take part in the collective motion of
the system. 
 
$R_{\rm AA}$ and $v_{2}$ are two complementary measurements to gain
insight into the heavy-quark transport coefficient of the
medium. Several theoretical model calculations are available for the
$R_{\rm AA}$  and  $v_{2}$ of heavy flavour hadrons.

\begin{figure}[t!]
\centering
\includegraphics[width=6cm, height=6cm]{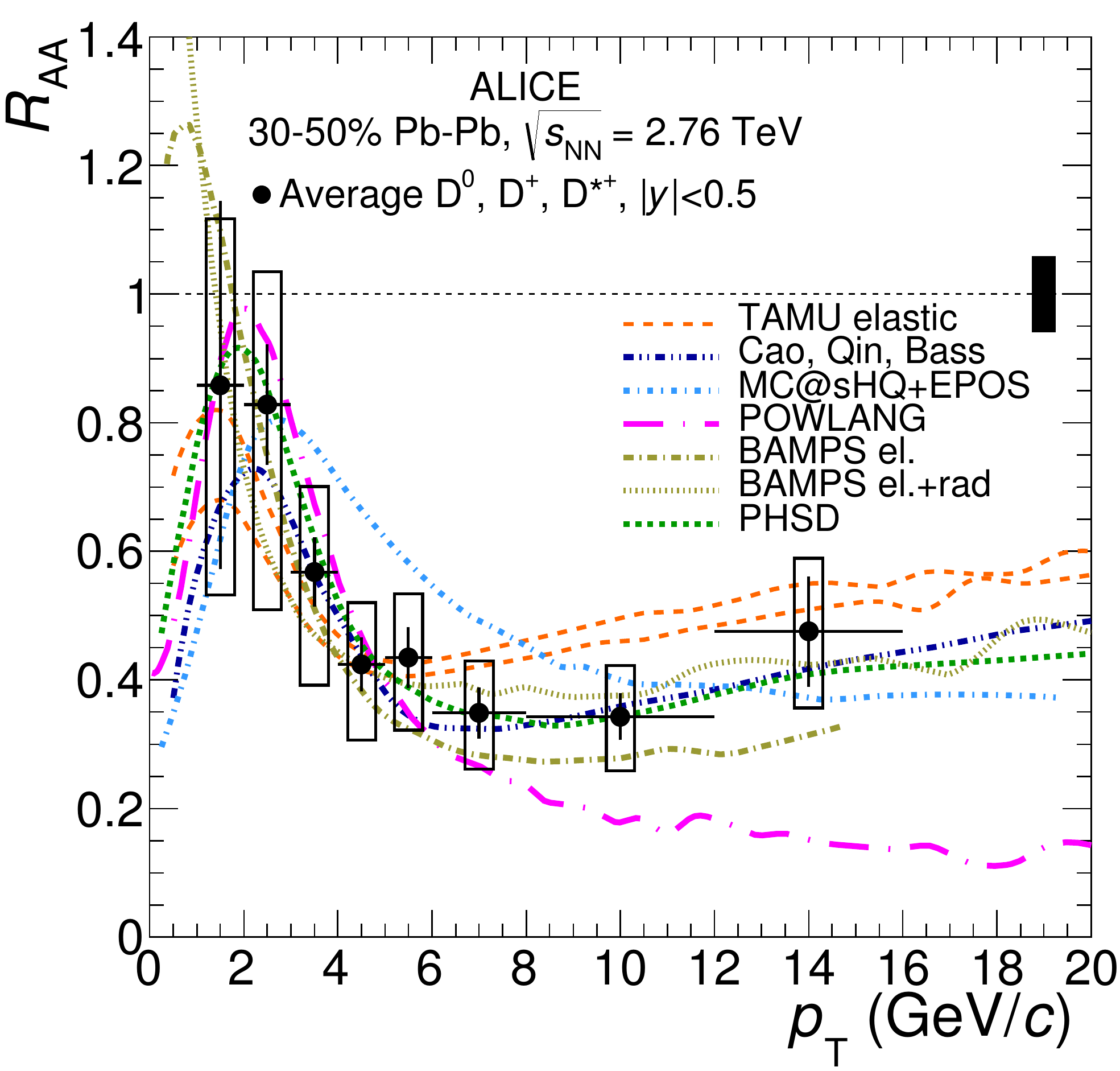}
\includegraphics[width=6cm, height=6cm]{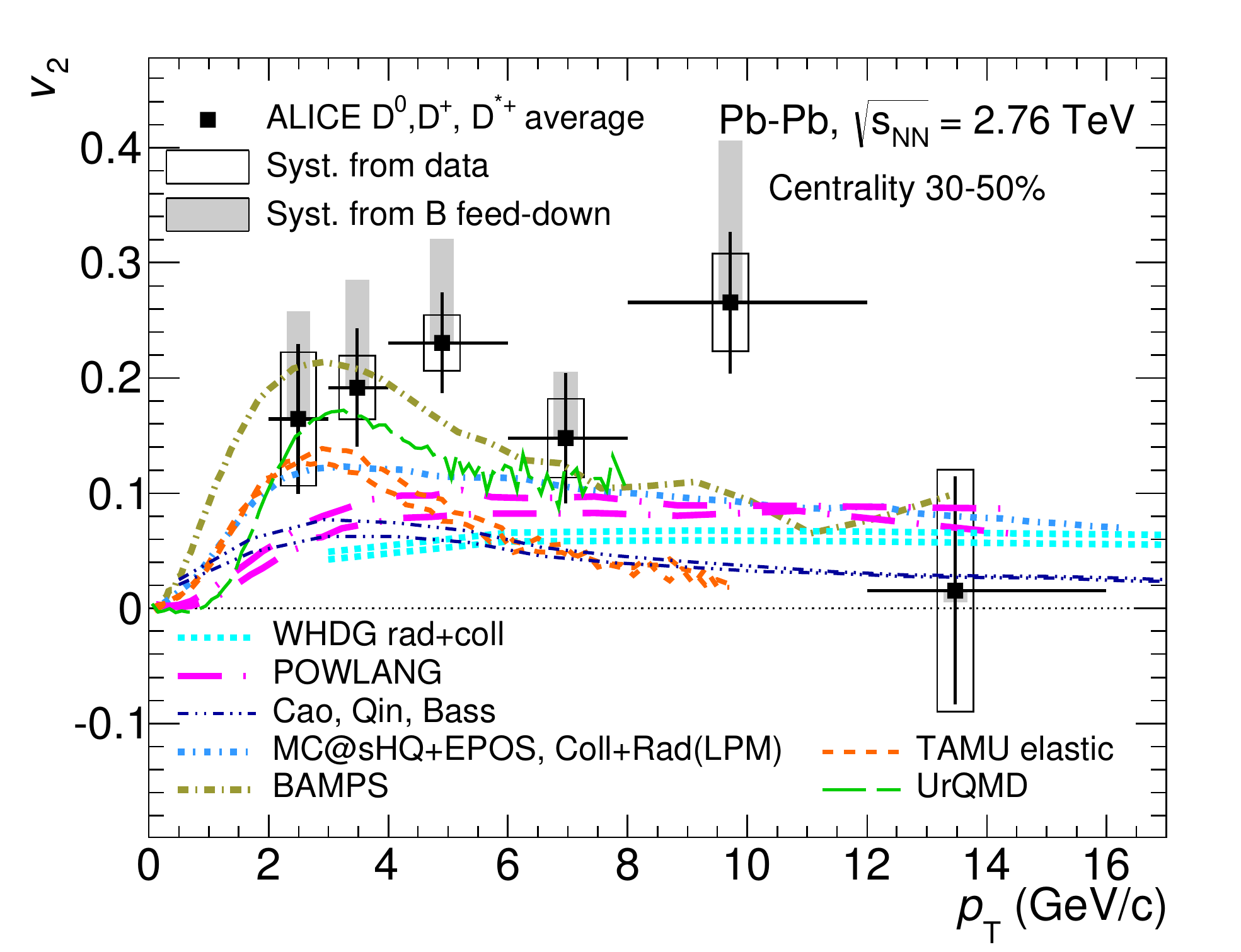}
\caption{(Color online). D meson $R_{\rm AA}$ and $v_{2}$ compared to model
predictions. Only models with predictions for both $R_{\rm AA}$ and
$v_{2}$ are shown. Left: D meson $R_{\rm AA}$ as a function of \pt. Right: D meson $v_{2}$ as a function of \pt.}
\label{Raav2model}
\end{figure}

Figure ~\ref{Raav2model} shows the D meson $R_{\rm AA}$ (left) and $v_{2}$ (right) compared to predictions
from various models~\cite{He:2014cla,Cao:2013ita,Nahrgang:2013xaa,Alberico:2011zy,Gomes:2013qza,Uphoff:2011ad,Fochler:2011en,Uphoff:2012gb}.
A simultaneous description
of the $R_{\rm AA}$ and $v_{2}$ starts to provide constraints to the models themselves.

\begin{figure}[t!]
\begin{center}
\includegraphics[width=6cm, height=6cm]{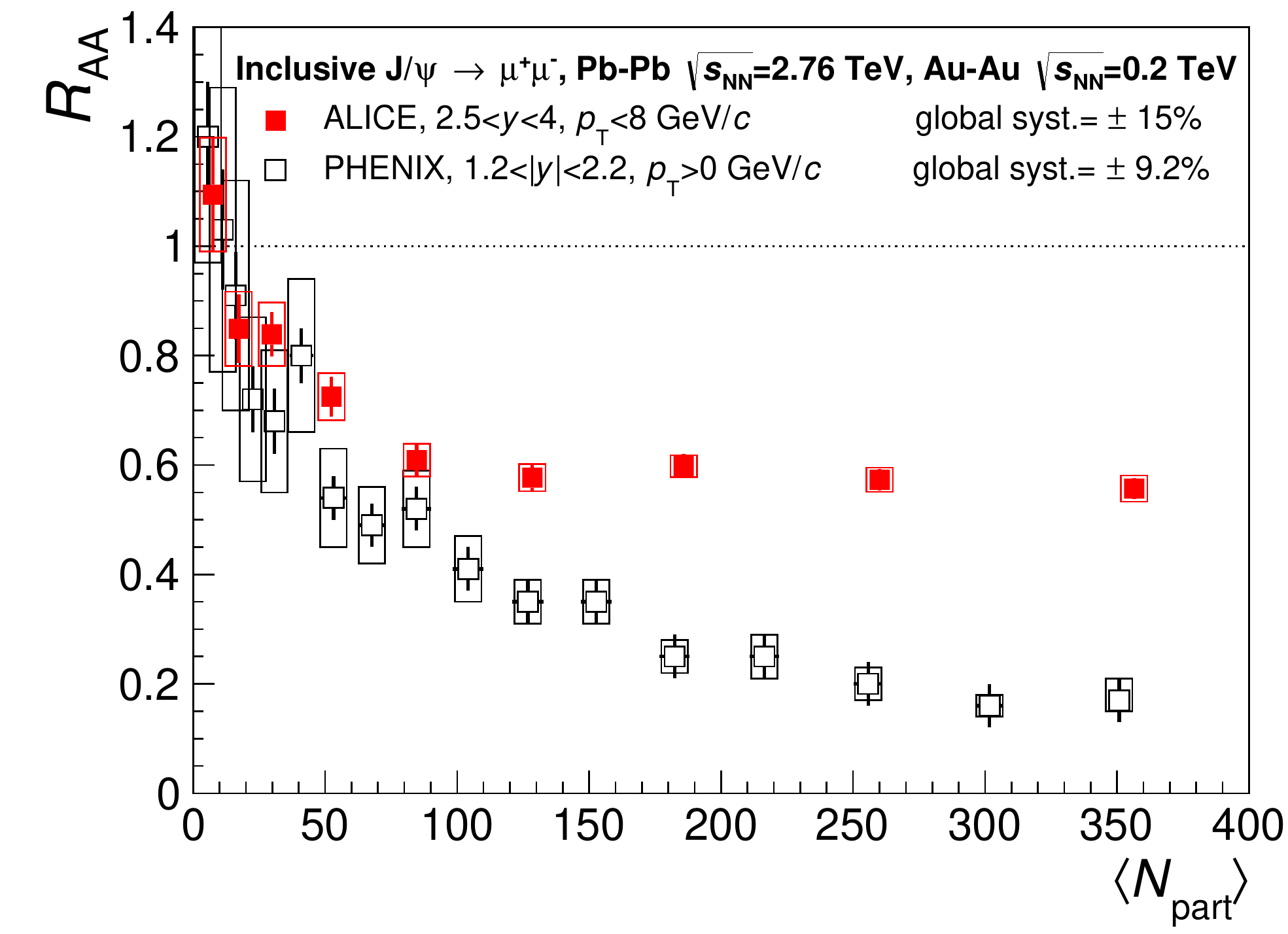}
\includegraphics[width=6cm, height=6cm]{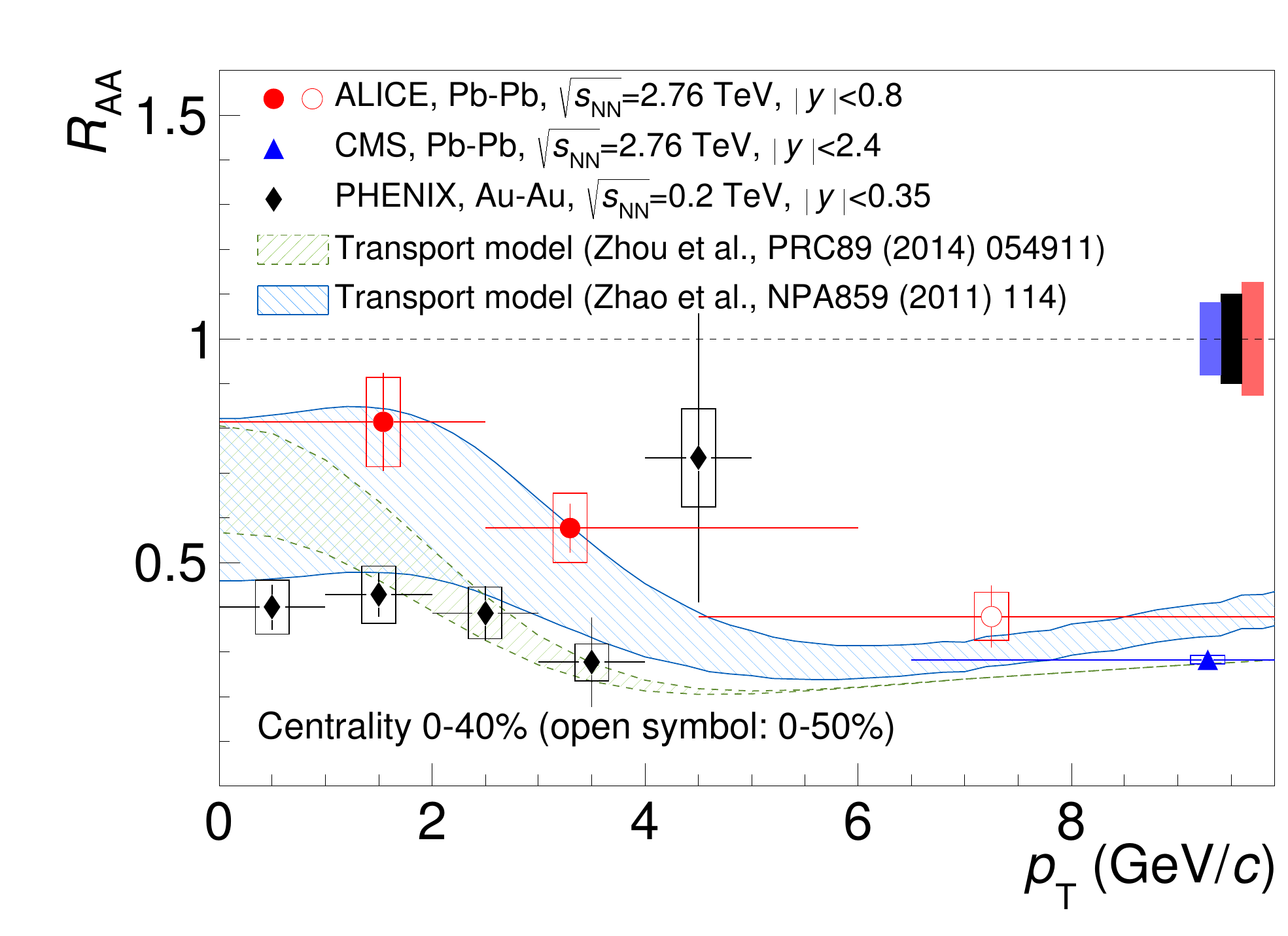}
\caption{(Color online). Left: Inclusive forward rapidity J/$\psi$ nuclear modification factor in \pbpb collisions at \snnt{2.76}~\cite{Abelev:2012rv}
and in \auau collisions at \snnt{0.2}~\cite{Adare:2006ns}. Right: $R_{\rm AA}$ as function of
transverse momentum  compared with measurements from CMS~\cite{Chatrchyan:2012np}  and PHENIX~\cite{Adare:2006ns} and
with theoretical models~\cite{Zhao:2011cv}}
\label{raajpsi}
\end{center}
\end{figure}

Figure~\ref{raajpsi} shows the J/$\psi$ $R_{\rm AA}$ measured in \pbpb collisions at  $\snnt{2.76}$  at forward rapidity as a function of the average number of participating nucleons $\langle N_{\rm part} \rangle$.
The ALICE results~\cite{Abelev:2012rv} are compared to measurements by PHENIX at RHIC in \auau collisions at $\snnt{0.2}$~\cite{Adare:2011yf}. The results for central collisions are significantly different with the ALICE data showing a factor of three less suppression than the PHENIX results. In addition, the RHIC data indicate a stronger centrality dependence, with the suppression increasing with centrality, whereas the LHC data are compatible with no centrality dependence. The similar behaviour is observed at mid-rapidity~\cite{Abelev:2013ila}. The \pt dependence of the J/$\psi$  $R_{\rm AA}$ at mid-rapidity for the 0-40\% most central collisions is shown in Fig.~\ref{raajpsi} (right). The data is compared to the PHENIX results~\cite{Adare:2006ns}, CMS results at higher \pt~\cite{Chatrchyan:2012np} and model calculations~\cite{Zhou:2014kka,Zhao:2011cv}. The ALICE data suggest a much lower degree of suppression at low \pt compared to lower energy measurements which is well described by transport model predictions~\cite{Zhou:2014kka}. Newer predictions~\cite{Zhao:2011cv} are systematically below the measurement and exhibit a \pt dependence similar to the one in the data. The model calculations show substantial theoretical uncertainties due to the limited knowledge of charm cross section and cold nuclear matter effects at LHC energies.

\section{Fluid-like behavior in small systems: \pp and \ppb collisions}
\label{sec:smallsystems}

To disentangle the so-called cold nuclear matter effects from those attributed to QGP, data from control experiments like \pp and \ppb collisions are analysed as a baseline. But surprisingly, results from the LHC showed that high multiplicity \pp and \ppb collisions exhibit characteristics reminiscent of those due to final state effects (flow-like patterns and the ridge structure), but no sign of jet quenching
~\cite{ABELEV:2013wsa,Abelev:2013haa} had been found so far. The possibility that QGP is formed in high multiplicity \pp events is not a new idea, for example, the FNAL-735 experiment at the Fermilab Tevatron collider had the main goal to search for evidence of a phase transition in ${\rm p\bar{p}}$ collisions at \sppt{1.8}~\cite{Friedlander:1979ci,Fowler:1986nv}. 

The understanding on the origin of these effects is ongoing, for the majority of the heavy-ion community they are interpreted as the evidence of the QGP formation in small systems~\cite{Gutay:2015cba,Bautista:2015kwa}. Although, another faction of the community is exploring new explanations in terms of initial state effects~\cite{Dusling:2012iga,Dusling:2012cg,Dusling:2015rja}, i. e., without invoking the formation of a small drop of QGP.

\begin{figure}[t!]
\begin{center}
   \includegraphics[width=0.95\textwidth]{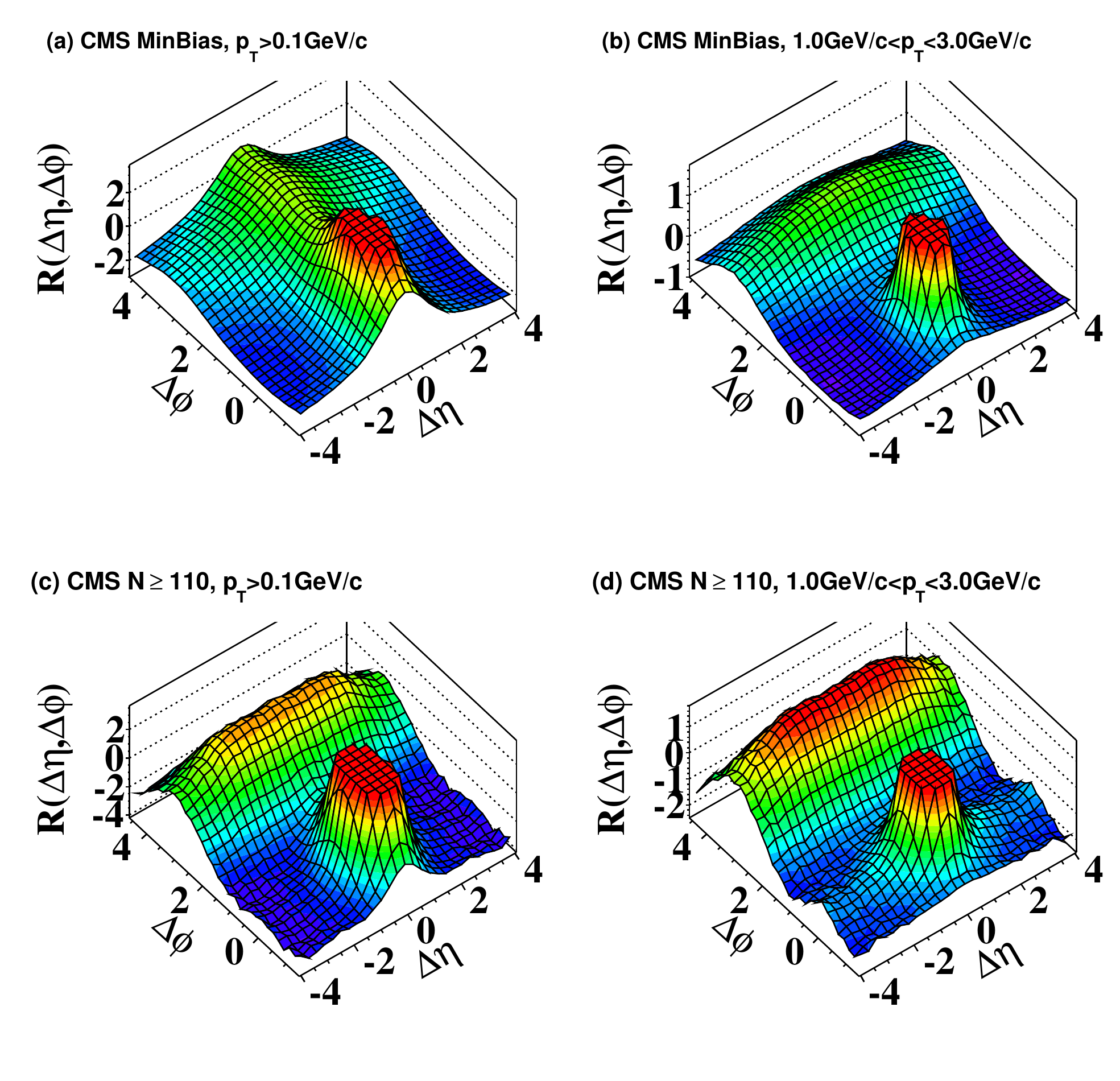}  
   \caption{(Color online). $\Delta\eta$-$\Delta\phi$ correlations for \pp collisions at \sppt{7} measured in (a) minimum bias events with \pt~$>$~0.1\,GeV/$c$, (b) minimum bias events with 1~$<$~\pt~$<$~3\,GeV/$c$, (c) high multiplicity events ($N_{\rm trk}^{\rm offline}$~$\geq$~110) with  \pt~$>$~0.1\,GeV/$c$ and (d) high multiplicity events with 1~$<$~\pt~$<$~3\,GeV/$c$.}
  \label{fig:pPb:1a}
\end{center}
\end{figure}

\subsection{The ridge structure in small systems}

Two-particle angular correlations provide important information about the hot and dense QCD matter formed in heavy-ion collisions. In particular at RHIC energies, long-range angular correlations (ridge structure) were reported~\cite{Alver:2008aa}, and they were attributed to the hydrodynamical evolution of the system. Motivated by the high particle densities produced in the highest multiplicity \pp collisions at the LHC energies, which were very close to those measured in high energy \cucu collisions, the CMS Collaboration studied the two-particle angular correlations of charged particles in \pp collisions at \sppt{7} and discovered the ridge structure in small systems~\cite{Khachatryan:2010gv}. Figure~\ref{fig:pPb:1a} shows the two-particle correlation functions measured in minimum bias and in high multiplicity \pp collisions. While the \pt-integrated correlation does not show any special feature, in the \pt range 1--3\,GeV/$c$ the near side long range angular correlation is clearly observed. Similar structures were also observed in \ppb collisions at \snnt{5.02}~\cite{CMS:2012qk,Abelev:2012ola,Aad:2012gla}. Furthermore, in high multiplicity events, non-zero second-order Fourier coefficients were extracted from the long-range correlations. Using the ALICE capabilities for particle identification,  the proton $v_{2}$ was observed to be smaller than that for pions, up to about \pt~=~2\,GeV/$c$~\cite{ABELEV:2013wsa}. This effect is similar to the mass ordering of $v_{2}$ previously discussed for heavy-ion collisions.

\begin{figure}[t!]
\begin{center}
   \includegraphics[width=0.95\textwidth]{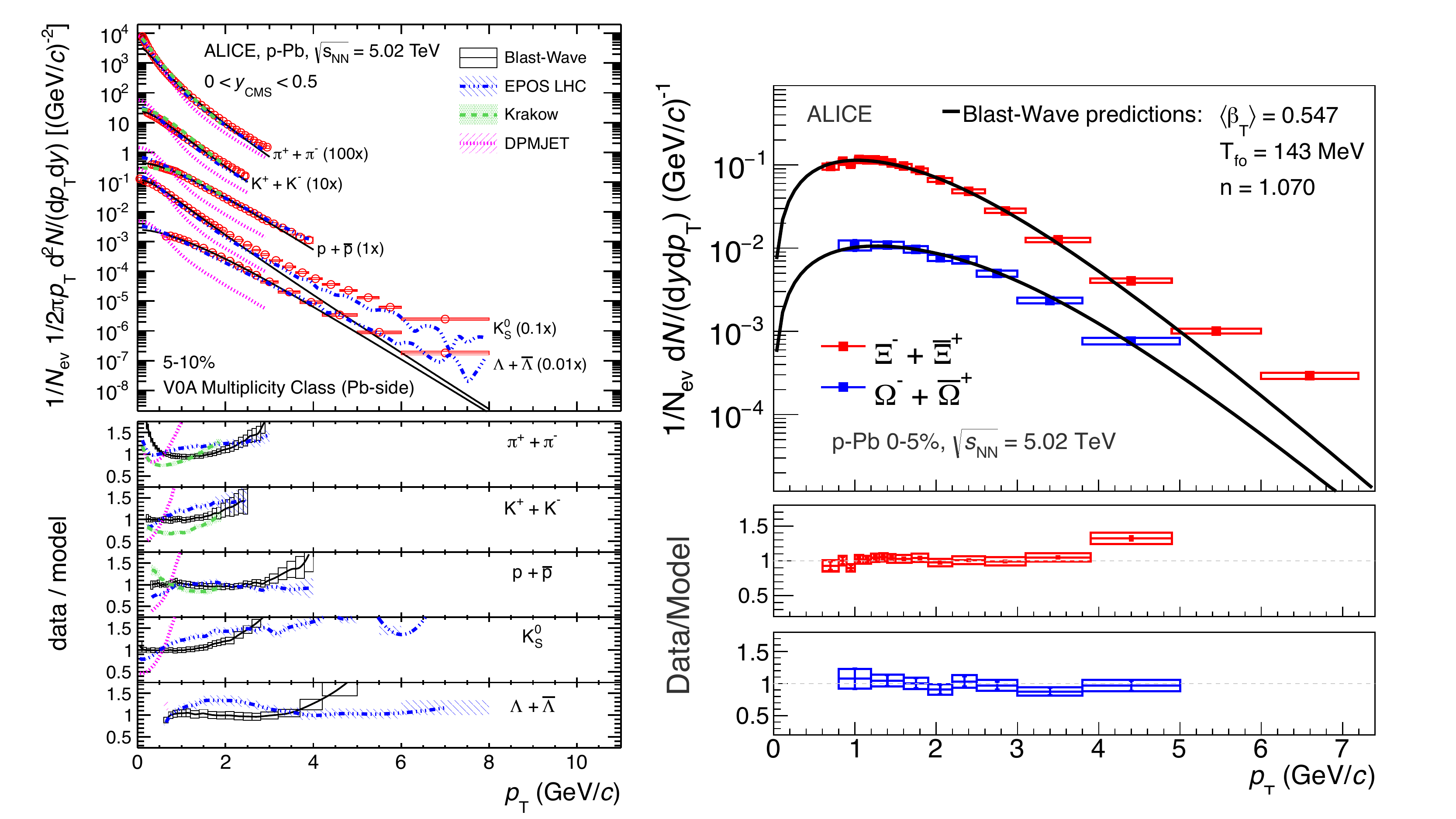}  
   \caption{(Color online).  Transverse momentum distributions of different particle species in the highest multiplicity class  are compared to the blast wave prediction obtained from the simultaneous blast-wave fit to the pion, kaon, proton and lambda \pt spectra.}
  \label{fig:pPb:1}
\end{center}
\end{figure}

\subsection{Light flavor production as a function of multiplicity in \ppb collisions}

The transverse momentum spectra of charged pions, kaons and (anti)protons as a function of the event multiplicity have been measured up to 20\,GeV/$c$~\cite{Adam:2016dau}. At low \pt ($<$~$2$--$3$\,GeV/$c$) the spectra exhibit a hardening with increasing multiplicity, with this effect being more pronounced for heavy particles. We are therefore observing features which resemble the radial flow effects well known from heavy-ion collisions~\cite{Adam:2015kca} and which are well described when a hydrodynamical evolution of the system is considered. At the LHC~\cite{Abelev:2013haa}  it was shown that for high multiplicity \ppb events, the \pt spectra were described by the blast-wave function. Using the parameters obtained from the simultaneous fit to pion, kaon, proton and lambda \pt spectra the model is able to describe the multi-strange baryon \pt distributions ($\pt$~$<$~4\,GeV/$c$)~\cite{Adam:2015vsf} as shown in Fig.~\ref{fig:pPb:1}. The feature is also observed in \pp collisions simulated with PYTHIA 8~\cite{Corke:2010yf,Ortiz:2013yxa}, where no hydrodynamical evolution is included, instead multiple partonic interactions (MPI) and color reconnection are producing the effects. The understanding of the role of MPI in data is ongoing by means of the so-called mini-jet analysis~\cite{Abelev:2013sqa,Abelev:2014mva} and event shapes~\cite{Abelev:2012sk,Cuautle:2015kra}.

\begin{figure}[t!]
\begin{center}
   \includegraphics[width=0.45\textwidth]{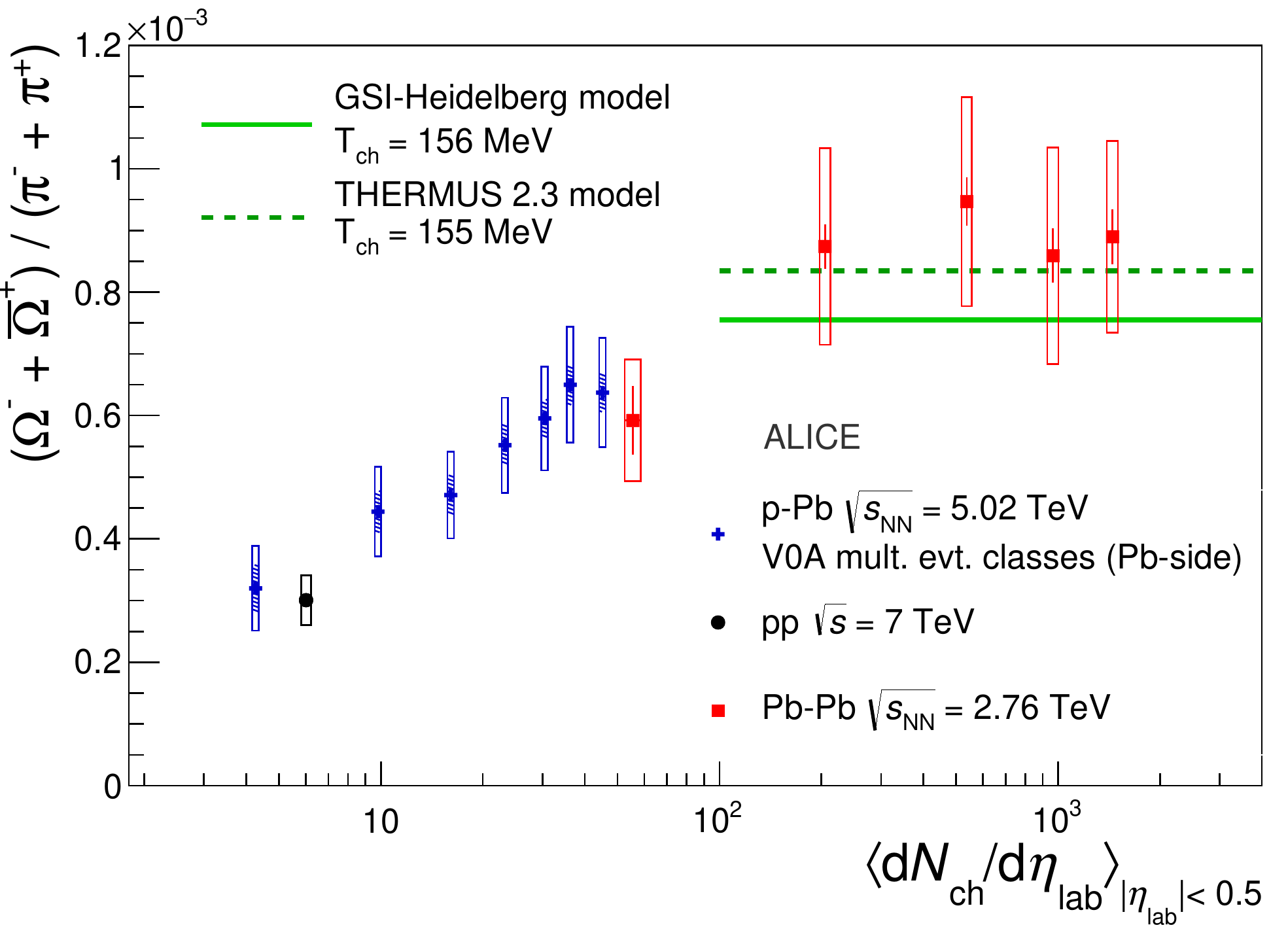} 
   \includegraphics[width=0.45\textwidth]{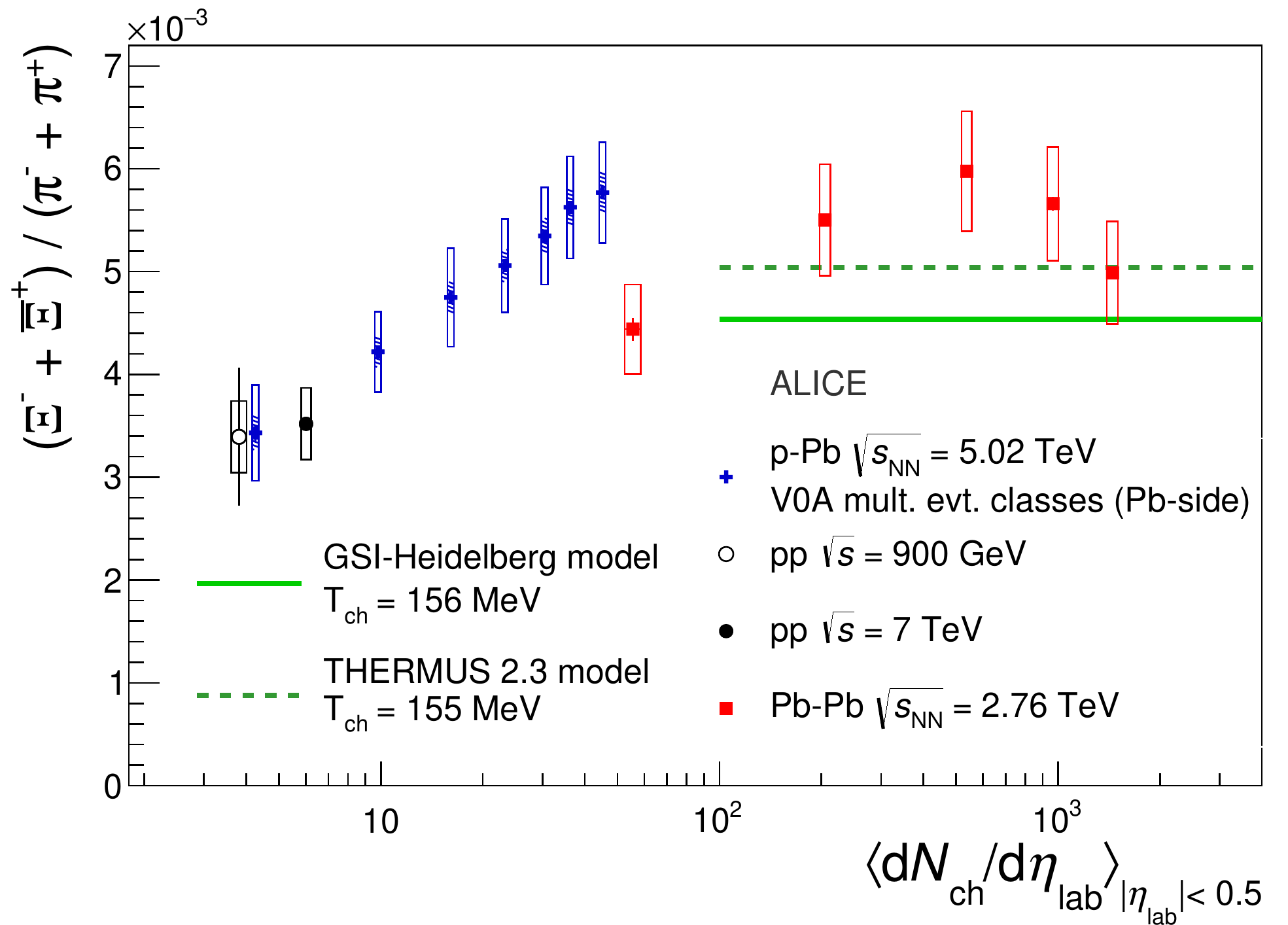}     
   \caption{(Color online).  $( \Xi^{-} + \bar{\Xi}^{+} ) / ( \pi^{+} + \pi^{-} )$ (left) and $( \Omega^{-} + \bar{\Omega}^{+} ) / ( \pi^{+} + \pi^{-} )$ (right) ratios as a function of the event multiplicity. Results for \ppb collisions at $\sqrt{s_{\rm NN}}=$5.02\,TeV are compared with minimum bias \pp ($\sqrt{s}=$0.9 and 7\,TeV) and \pbpb results.}
  \label{fig:pPb:3}
\end{center}
\end{figure}

To study the relative production of strangeness and compare it with results in minimum-bias \pp and \pbpb collisions, the ratios to charged pions have been measured as a function of the event multiplicity~\cite{Adam:2015vsf}. Figure~\ref{fig:pPb:3} shows that both the $( \Xi^{-} + \bar{\Xi}^{+} ) / ( \pi^{+} + \pi^{-} )$ and $( \Omega^{-} + \bar{\Omega}^{+} ) / ( \pi^{+} + \pi^{-} )$ ratios increase as a function of the event multiplicity. The relative increase is more pronounced for $\Omega$ than for $\Xi$, these relative increases are larger than the 30\% increase observed for $\Lambda/\pi$ ratio, suggesting that strangeness content may control the rate of increase with multiplicity. Also interesting is the fact that the maximum increases reach the values measured for \pbpb collisions, and that the $\Xi/\pi$ ratio is systematically above the thermal model predictions. In the context of heavy-ion collisions, this effect (strangeness enhancement) has been considered a signature of the QGP formation. Recently, it has been pointed out that a perfect scaling of the particle ratios with the energy density holds for the different colliding systems~\cite{guy:focus,cuautle:energy} opening new possibilities for a better understanding of the QGP-like features in small systems.

\begin{figure}[t!]
\begin{center}
   \includegraphics[width=0.55\textwidth]{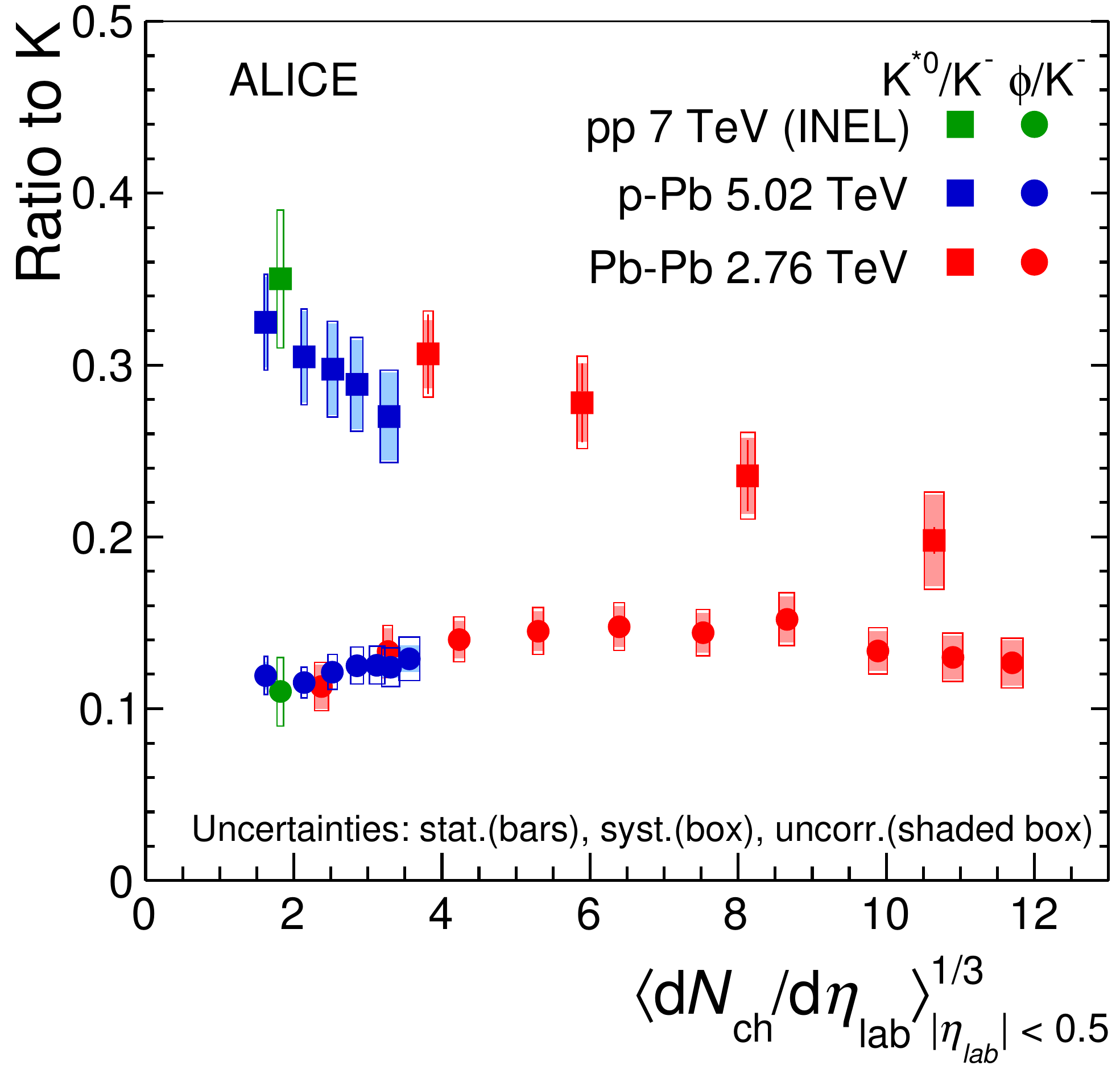}   
   \caption{(Color online).  K$^{*0}$ and $\phi$ yields normalized to that for charged kaons as a function of the fireball size, results for different colliding systems are shown: \pbpb, \ppb and  minimum bias \pp collisions.}
  \label{fig:pPb:4}
\end{center}
\end{figure}

In order to look for the presence of re-scattering effects in high multiplicity \ppb collisions; the K$^{*0}$ and $\phi$ relative to charged kaons production is studied as a function of the cube root of the average mid-rapidity charged particle density. In Fig.~\ref{fig:pPb:4}, a comparison with minimum bias \pp collisions at $\sppt{7}$ and \pbpb collisions at $\snnt{2.76}$ is presented. In heavy-ion collisions the decreasing trend of K$^{*0}$$/$K with increasing fireball size has been explained as a consequence of a re-scattering of K$^{*0}$ decay daughters in the hadronic phase. It is worth noticing that a similar trend is also observed in \ppb collisions.

\begin{figure}[t!]
\vspace{-0.5cm}
\begin{center}
   \includegraphics[width=0.85\textwidth]{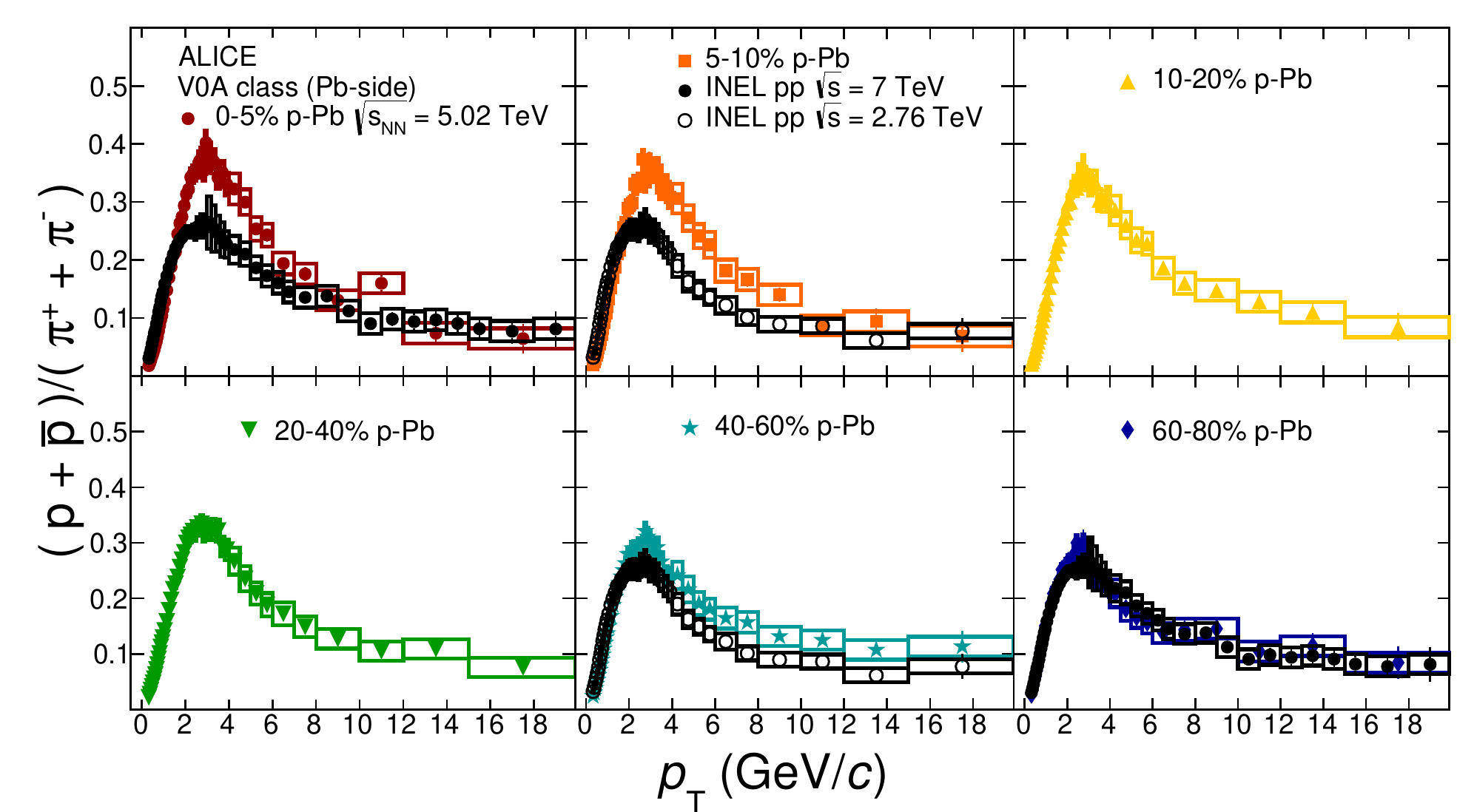}   
   \caption{(Color online).  Proton-to-pion ratios as a function of $p_{\rm T}$ for different multiplicity classes. Results for \ppb collisions (full markers) are compared to the ratios measured in INEL pp collisions at 2.76 TeV~\cite{Abelev:2014laa} (empty circles) and at 7 TeV (full circles).}
  \label{fig:pPb:5}
\end{center}
\end{figure}

The multiplicity dependence of the intermediate to high-\pt particle production is studied with the particle ratios~\cite{Adam:2016dau}. The proton-to-pion ratio as a function of the event multiplicity is shown in Fig.~\ref{fig:pPb:5}. The particle ratios exhibit a maximum (bump) at \pt~$\approx$~3\,GeV/$c$ and the size of the bump increases with increasing multiplicity. On the other hand, at higher transverse momenta (\pt$>$~10\,GeV/$c$) the ratios return to the values measured for \pp and \pbpb collisions. Any particle species dependence of the nuclear modification factor is therefore excluded.

\subsection{Heavy flavors}

\begin{figure}[t!]
\includegraphics[width=6cm, height=6cm]{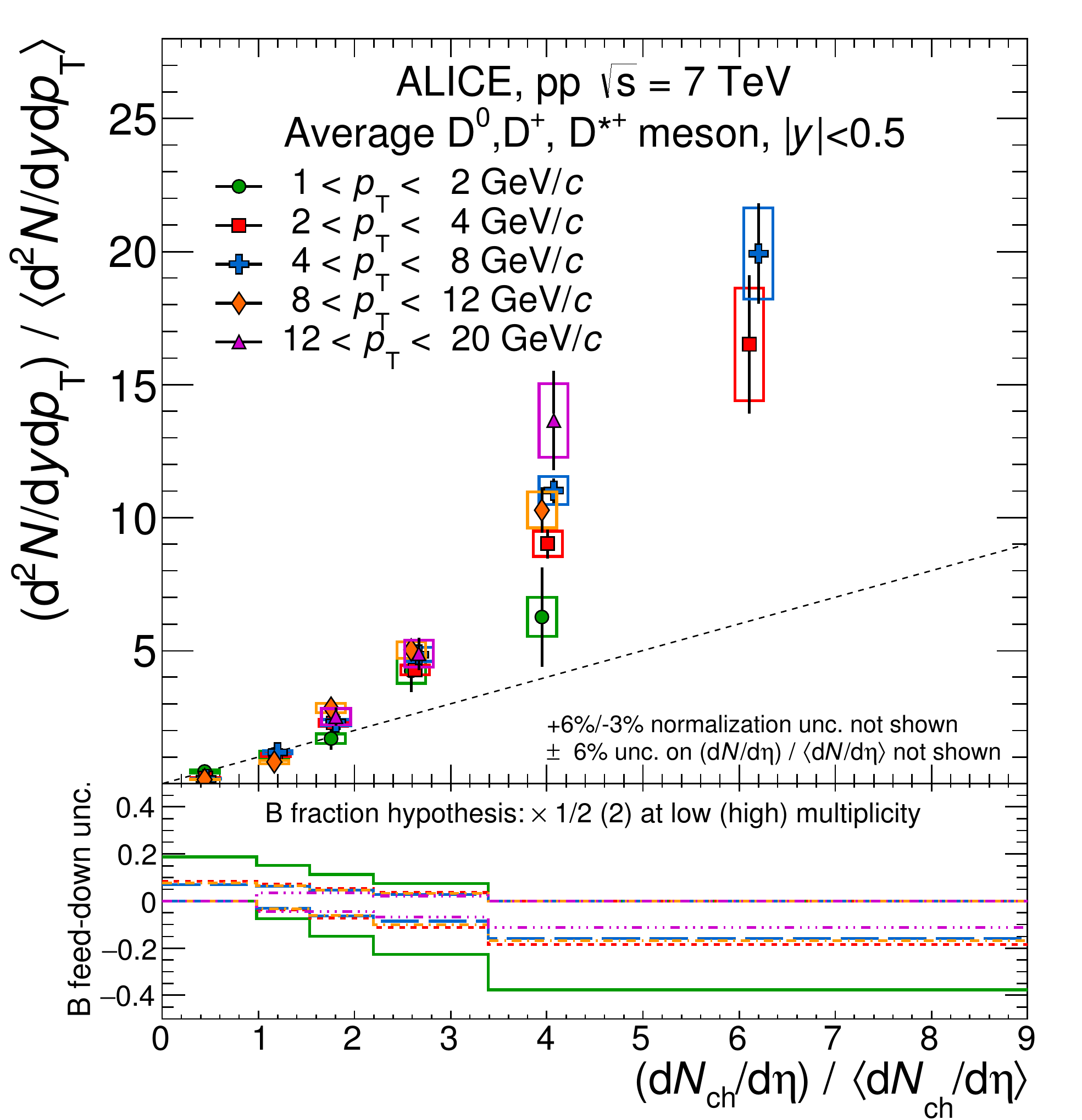}
\includegraphics[width=6cm, height=6cm]{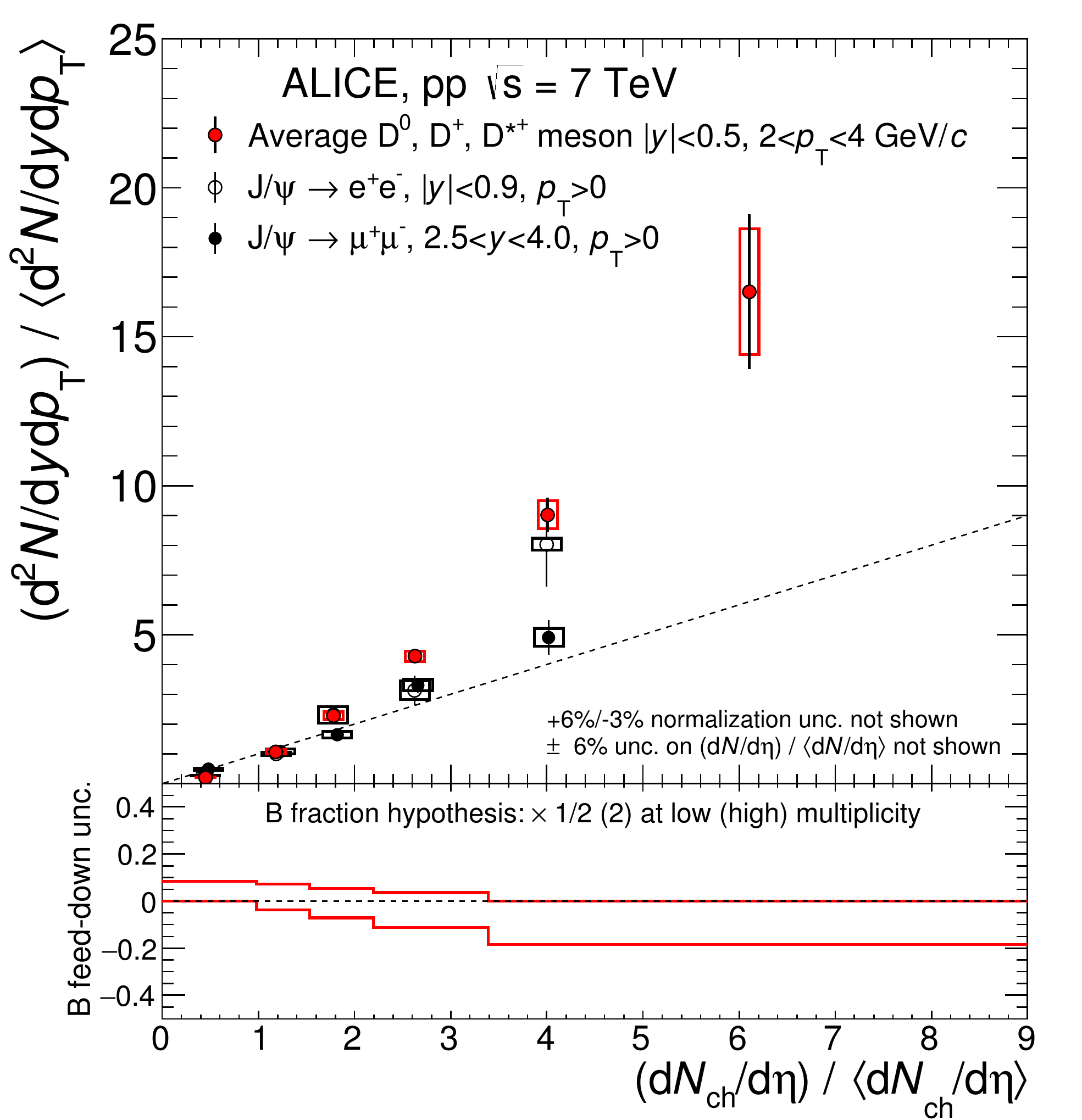}
\caption{(Color online). Self normalized average D meson yields as a function of multiplicity in pp collisions for different \pt ranges Left: multiplicity estimates at central rapidity. Right: D meson and J/$\psi$ yields as a function of charged particle multiplicity at central rapidity}
\label{3}
\end{figure}

The measurement of heavy-flavour production as a function of the multiplicity of charged particles produced in
hadronic collisions is sensitive to the interplay between hard and soft contributions to particle production and could
give insight into the role of MPI. \\
The study of the D meson production, evaluated for various multiplicity and \pt intervals, is presented via the D meson self normalized yield, i.e, the corrected per-event yield normalized to the multiplicity-integrated value. The results for $\rm D^{0}$, $\rm D^{+}$ and $\rm D^{*+}$ are compatible and the average self-normalized yield of the three species increases with increasing relative multiplicity at mid-rapidity and does not depend on the \pt within uncertainties in the interval 1--20\,GeV/$c$, as shown in Fig.~\ref{3} (left). A similar trend is observed when the multiplicity is estimated at forward rapidity, demonstrating that the trend is not connected to a possible bias due to the pseudorapidity region of the measurements.  The self normalized yield of D mesons is comparable with that of the inclusive J/$\psi$ measured at mid rapidity as well as of the non-prompt J/$\psi$. The similar increases with multiplicity for open charm, open beauty and charmonia at mid-rapidity suggest that this effect is not (or is slightly) due to the hadronization mechanism, but likely related to the heavy flavour production mechanisms. The yield increase can be described by calculations taking into account the contributions of MPI, by the influence of the interactions between colour sources in the percolation model, or by the effect of the initial conditions of the collision followed by a hydrodynamic evolution computed with the EPOS 3 event generator. More precise measurements are needed to discriminate among the possible origins of the effect.

\begin{figure}[t!]
\begin{center}
\includegraphics[width=6cm, height=6cm]{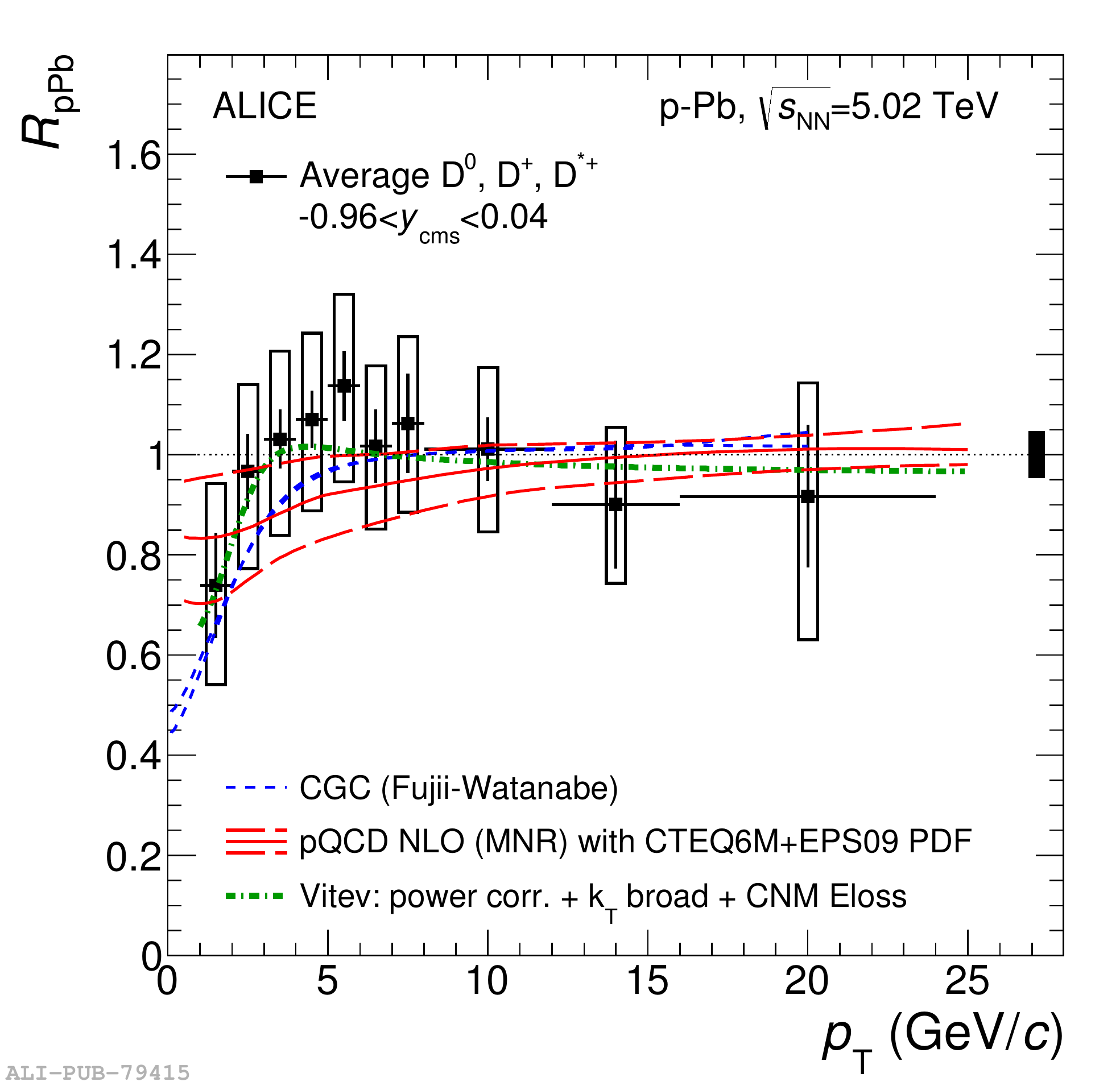} 
\includegraphics[width=6cm, height=6cm]{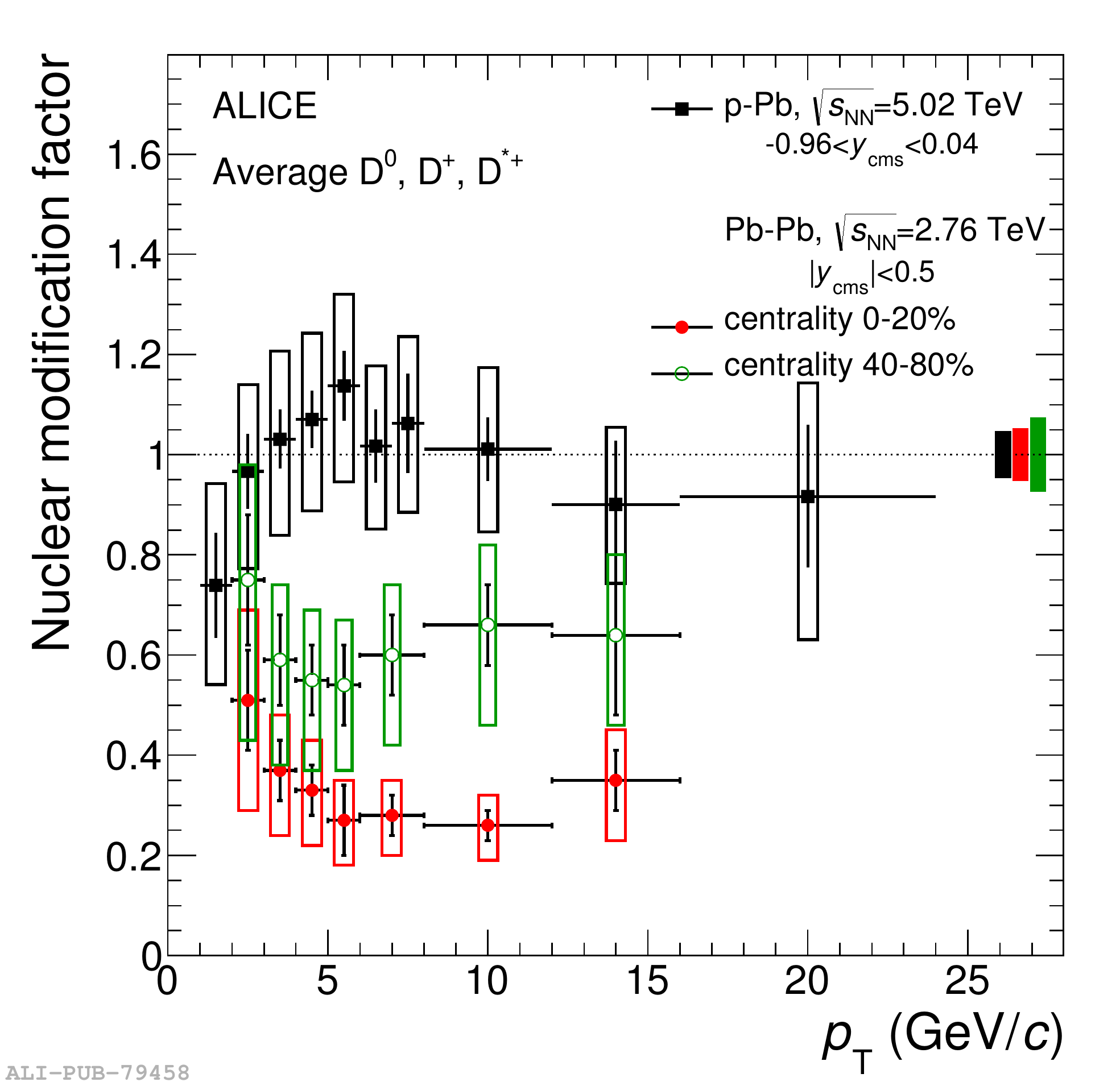}
\caption{(Color online). Average $R_{\rm pPb}$ of D mesons as a function of \pt compared to theoretical predictions (left) and the D meson \raa for the 0-20\% and 40-80\% \pbpb collisions (right).}
\label{fig4}
\end{center}
\end{figure}

In \ppb collisions CNM effects, such as the modification of the parton distribution functions due to the presence of the nucleus, are expected to affect the heavy-quark yield and \pt distributions relative to \pp collisions. In particular, by measuring heavy flavour hadron production in different \pt ranges, it is possible to access different Bjorken-$x$ regimes.
The \ppb spectra are quantitatively compared to the \pp reference by computing the nuclear modification factor, $R_{\rm pPb}$. The average $R_{\rm pPb}$ for D$^{0}$, D$^{+}$ and D$^{*+}$ in the \pt-range 1~$<$~\pt~$<$~24\,GeV/$c$ is shown in Fig.~\ref{fig4} (left) together with the comparison with theoretical calculations. It can be observed that the $R_{\rm pPb}$ of prompt D mesons is described within the uncertainties by different models including initial state effects ~\cite{Abelev:2014hha}. The measurements confirm that initial and final state effects due to the presence of CNM are small in the measured $p_{\rm T}$ range.
Figure~\ref{fig4} (right) shows the average D meson nuclear modification
factor measured in minimum bias \ppb collisions at \snnt{5.02}~\cite{Abelev:2014hha} compared to D meson $R_{\rm AA}$ measured in 0-20$\%$ and 40-80$\%$ \pbpb collisions. Since no significant modification of the
D meson yield is observed in \ppb collisions for \pt$>$~2\,GeV/$c$, it can be concluded that the strong suppression observed in
central \pbpb collisions is due to the interaction of heavy quarks with the hot and dense QCD medium. 

\begin{figure}[t!]
\begin{center}
\includegraphics[width=6cm, height=6cm]{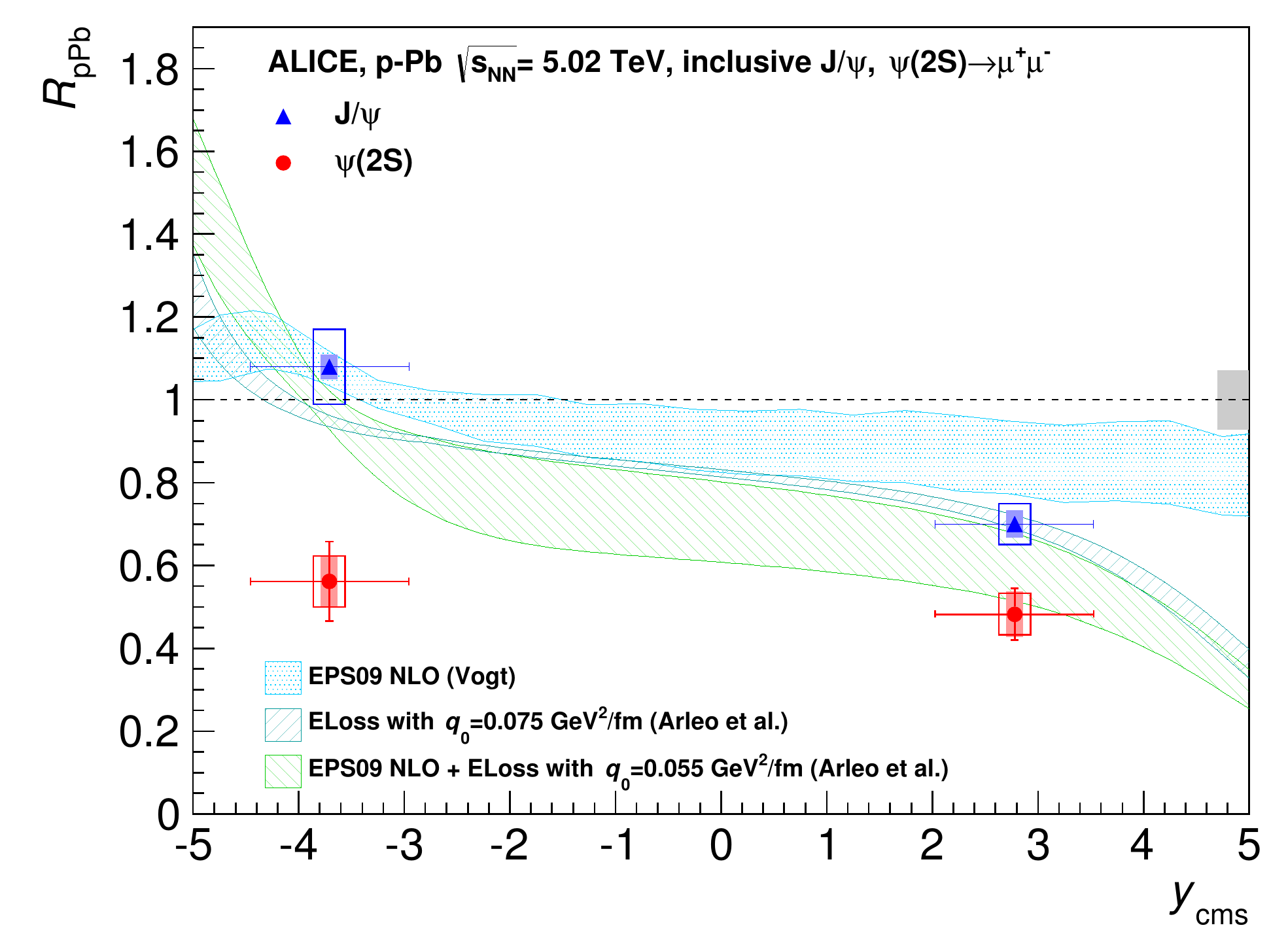}
\includegraphics[width=6cm, height=6cm]{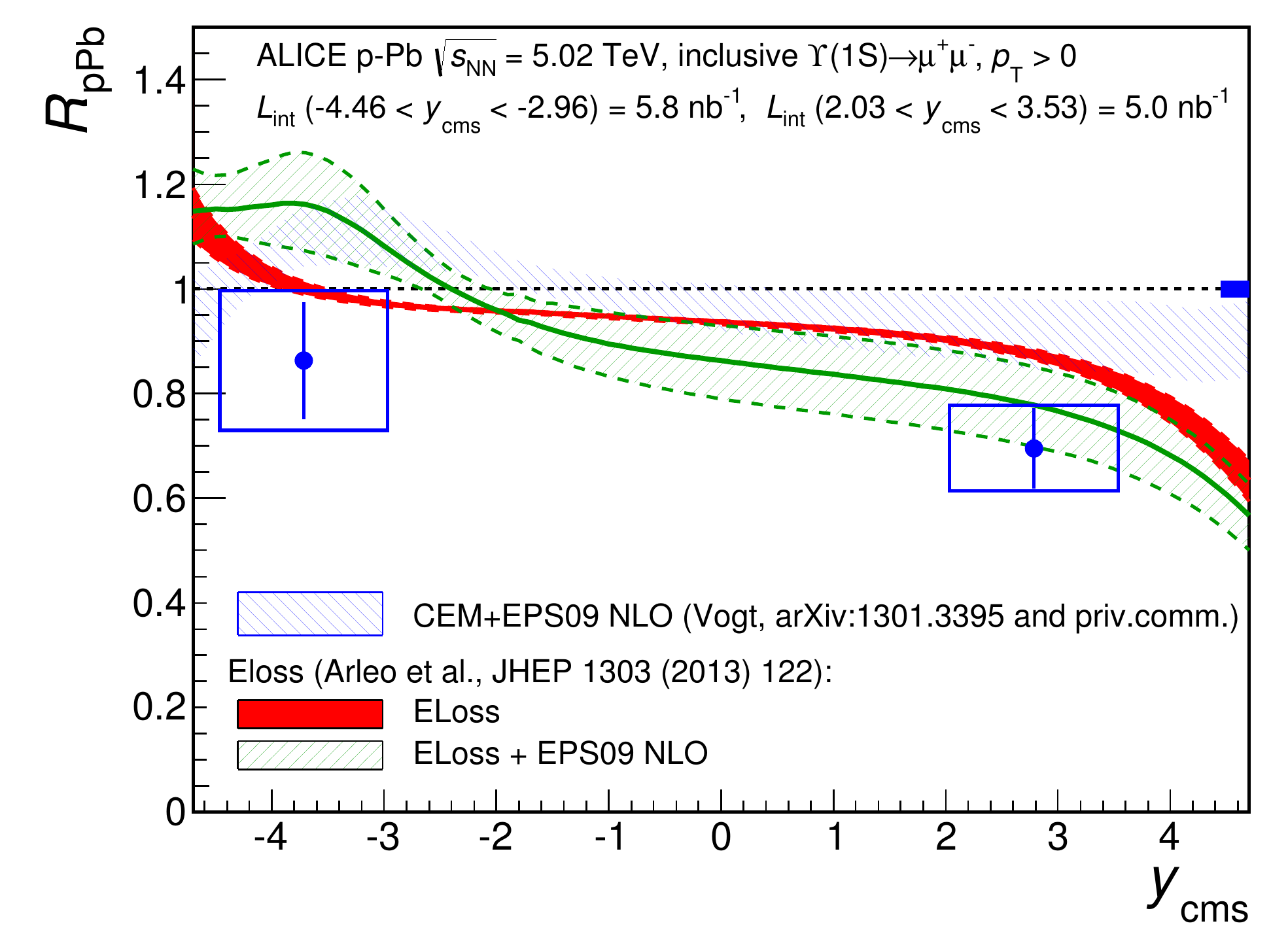}
\caption{(Color online). The $R_{\rm pPb}$ of inclusive J/$\psi$ and $\psi$(2S) as measured in backward and forward rapidity (left panel).
The $\Upsilon$(1S) $R_{\rm pPb}$ as measured in forward and backward rapidity (right panel).}
\label{jpsi2psi}
\end{center}
\end{figure}

The inclusive J/$\psi$, $\psi$(2S) and $\Upsilon$(1S) production in \ppb collisions has been also measured by ALICE at $\snnt{5.02}$~\cite{Abelev:2013yxa,Abelev:2014zpa,Abelev:2014oea}. In the absence of the \pp reference, an interpolation between lower energy ($\sppt{2.76}$) and higher energy ($\sppt{7}$) measurements has been used for the calculation of $R_{\rm pA}$~\cite{Abelev:2013yxa,Abelev:2014zpa,Abelev:2014oea}. The nuclear modification factor of J/$\psi$ and $\psi$ (2S)  are compared along with model predictions in the left panel of Fig.~\ref{jpsi2psi}. The models depending on shadowing and coherent parton energy loss with or without shadowing can explain the J/$\psi$ $R_{\rm pA}$~\cite{Arleo:2012hn,Vogt:2001ky}. Since no dependence on the quantum number of the resonances is considered, the model predictions for the $\psi$ (2S) are identical to the J/$\psi$ ones. Theoretical calculations based on shadowing and/or energy loss can not explain simultaneously the J/$\psi$ and the $\psi$(2S) behavior, in particular at backward rapidity (-4.46~$< y_{\rm cms} <$~-2.96). While model calculations are in good agreement with the J/$\psi$, they strongly underestimate the $\psi$(2S) suppression. The coherent energy loss model calculation including shadowing reproduces the $\Upsilon$ (1S) $R_{\rm pPb}$ at forward rapidity but overestimate it at backward rapidity. An opposite trend is found for the coherent energy loss calculation without shadowing (right panel of Fig. ~\ref{jpsi2psi}).

\section{Summary}

In this paper we have presented a selection of experimental results from the LHC heavy-ion Run I. In \pbpb collisions, the presented results significantly improve the precision of previous measurements in various areas. In particular, a measurement of elliptic flow with identified particles shows a clear mass ordering for light and strange hadrons for \pt~$<$~2.5\,GeV/$c$.  For \pt~$<$~4\,GeV/$c$, spectra and $v_{2}$ measurements of the meson suggest that the mass (and not the number of constituent quarks) drives the spectral shape and the size of the elliptic flow only in the 0-40\% central collisions. Contrary, for the more peripheral collisions, the number of constituent quarks starts to play a role. 

While there are several observables which are approximately consistent with a description  of  \ppb  collisions  as an  incoherent  superposition  of  nucleon-nucleon  collisions  at  high \pt,  some measurements hint to novel effects at low \pt which are potentially of collective origin. For example, we have discussed that for high multiplicity \ppb collisions; the \pt spectra are well described by hydrodynamical calculations,  experiments also found long range angular correlations and, a $v_{2}$ which is not zero and exhibits a mass ordering. Moreover, the so-called strangeness enhancement was reported too, namely, the hyperon to pion ratios increase with multiplicity from the values measured in minimum bias \pp to those observed in \pbpb collisions. However, no jet quenching signatures in small systems have been found so far, therefore, we still cannot rule out alternative explanations which do not invoke the QGP formation in small systems. Clearly, these findings still need to be reconciled theoretically and promise that \pp and \ppb  collisions will continue to be a very hot topic in the future.

\section*{Acknowledgements}

The authors acknowledge Guy Pai\'c for the critical reading of the manuscript and the valuable discussion and suggestions. Support  for  this  work  has  been  received  from  CONACYT  under  the  grant  No.   260440;  from
DGAPA-UNAM under PAPIIT grant IA102515 and also from Council of Scientific and Industrial Research (CSIR) New Delhi and from the grant LG 15052 of the Ministry of Education of the Czech 
Republic. The EPLANET program supported the mobility between UNAM and CERN.

\bibliographystyle{ws-ijmpe}
\bibliography{biblio_final}

\end{document}